\documentclass[fleqn,usenatbib,useAMS]{mnras}
\usepackage{newtxtext,newtxmath}
\usepackage[T1]{fontenc}
\DeclareRobustCommand{\VAN}[3]{#2}
\let\VANthebibliography\thebibliography
\def\thebibliography{\DeclareRobustCommand{\VAN}[3]{##3}\VANthebibliography}

\usepackage{graphicx}	% Including figure files
\usepackage{tabularx}
\usepackage{authblk}
\usepackage{orcidlink}
\usepackage{color}
\usepackage{xcolor}

\title[SKA Science Data Challenge 3a]{Square Kilometre Array Science Data Challenge 3a: foreground removal for an EoR experiment}
\author{ A.  Bonaldi \orcidlink{0000-0003-4787-2888}$^{1}$,  P.  Hartley$^{1}$,  R.  Braun$^{1}$,  S.  Purser \orcidlink{0000-0002-6358-5152}$^{1}$,  A.  Acharya \orcidlink{0000-0003-3401-4884}$^{2}$,  K.  Ahn \orcidlink{0000-0003-3974-1239}$^{3}$,  M.  Aparicio Resco \orcidlink{0000-0001-5094-2731}$^{4}$,  O.  Bait \orcidlink{0000-0003-2722-8841}$^{1}$,  M.  Bianco \orcidlink{0000-0002-6766-0017}$^{5}$,  A.  Chakraborty \orcidlink{0000-0002-7758-9859}$^{6}$
% ...
\newline
\it{Affiliations at the end of the paper}}
\newpage

\begin{document}
\date{Accepted XXX. Received XXX; in original form XXX}

%\pagerange{\pageref{firstpage}--\pageref{lastpage}} \pubyear{2017}

\maketitle

\clearpage
\begin{abstract}
We present and analyse the results of the Science data challenge 3a (SDC3a, \url{https://sdc3.skao.int/challenges/foregrounds}), an EoR foreground-removal community-wide exercise organised by the Square Kilometre Array Observatory (SKAO). The challenge ran for 8 months, from March to October 2023. Participants were provided with realistic simulations of SKA-Low data between 106\,MHz and 196\,MHz, including foreground contamination from extragalactic as well as Galactic emission, instrumental and systematic effects. They were asked to deliver cylindrical power spectra of the EoR signal, cleaned from all corruptions, and the corresponding confidence levels. Here we describe the approaches taken by the 17 teams that completed the challenge, and we assess their performance using different metrics. 

The challenge results provide a positive outlook on the capabilities of current foreground-mitigation approaches to recover the faint EoR signal from SKA-Low observations. The median error committed in the EoR power spectrum recovery is below the true signal for seven teams, although in some cases there are some significant outliers. The smallest residual overall is $4.2_{-4.2}^{+20} \times 10^{-4}\,\rm{K}^2h^{-3}$cMpc$^{3}$ %{\bf add percentage values} 
across all considered scales and frequencies. 

The estimation of confidence levels provided by the teams is overall less accurate,  with the true error being typically under-estimated, sometimes very significantly. The most accurate error bars account for $60 \pm 20$\% of the true errors committed. The challenge results provide a means for all teams to understand and improve their performance. This challenge indicates that the comparison between independent pipelines could be a powerful tool to assess residual biases and improve error estimation. 

\end{abstract}
%\twocolumn 

\section{Introduction}
The redshifted 21\,cm signal, produced by the hyperfine spin-flip transition of neutral hydrogen, is a powerful tool for mapping the distribution of neutral gas during the cosmic evolution that followed recombination \citep{field58, madau97}. Studying this signal provides valuable insights into the astrophysical processes shaping three pivotal stages of cosmic history: the cosmic dawn (CD), the epoch of reionization (EoR) and the post-reionization era. 

During the CD, neutral gas began to accumulate within dark matter halos. As these halos grew dense enough to undergo gravitational collapse, they formed the first luminous objects in the Universe. In the subsequent stage, the EoR, radiation from the first stars, galaxies, and quasars gradually ionized the surrounding neutral gas, transforming the Universe from predominantly neutral to mostly ionized. Finally, the Universe entered the post-reionization era, during which the intergalactic medium remained ionized, with only traces of neutral hydrogen confined to dense regions like galaxies.

Efforts to measure the redshifted 21\,cm signal have led to the development of a variety of low-frequency radio facilities, which include both single-antenna instruments and interferometers. Single-antenna instruments aim to measure the globally averaged redshifted 21\,cm signal across the sky, providing insights into the overall evolution of neutral hydrogen during different cosmic stages. Interferometers are designed to detect spatial variations in the redshifted 21\,cm signal, enabling detailed mapping of the distribution of neutral hydrogen and ionized regions through cosmic history. These facilities, however, face several challenges, including the need to mitigate foreground contamination from Galactic synchrotron emission and extragalactic sources \citep{shaver99, jelic08, chapman19}, as well as terrestrial radio frequency interference. Additionally, achieving precise instrumental calibration is essential, as any systematic errors can mask the faint redshifted 21\,cm signal and compromise its accurate measurement.

Despite the challenges, significant progress has been made over the last two decades in placing upper limits on the redshifted 21\,cm signal power spectrum from the EoR using existing low-frequency interferometers. Instruments such as the Giant Metrewave Radio Telescope \citep[GMRT;][]{paciga11}, the Precision Array to Probe the EoR \citep[PAPER;][]{kolopanis19}, the LOw-Frequency ARray \citep[LOFAR;][]{patil17, mertens20}, the Murchison Widefield Array \citep[MWA;][]{barry19, li19, trott20}, and the Hydrogen Epoch of Reionization Array \citep[HERA;][]{abdurashidova22} have progressively tightened these constraints. Furthermore, several facilities, including the MWA \citep{ewallwice16, yoshiura21}, the Owens Valley Radio Observatory – Long Wavelength Array \citep[OVRO-LWA;][]{eastwood19, garsden21}, the LOFAR Amsterdam ASTRON Radio Transients Facility and Analysis Center \citep[LOFAR AARTFAAC;][]{gehlot20}, and the New Extension in Nançay Upgrading LOFAR \citep[NenuFAR;][]{munshi24}, are also beginning to place constraints on the 21\,cm signal from the CD. These efforts are complemented by results from single-antenna experiments, such as the Experiment to Detect the Global Epoch of Reionization Signature \citep[EDGES][]{bowman08, bowman18} and the Shaped Antenna measurement of the background RAdio Spectrum \citep[SARAS;][]{singh17, nambissan21}.

The new generation of interferometers, such as the Hydrogen Epoch of Reionization Array \citep[HERA;][]{deboer17} and the Square Kilometre Array - Low \citep[SKA-Low;][]{koopmans15}, is poised to significantly advance the search for the redshifted 21\,cm signal. These instruments, with their enhanced sensitivity, are designed to enable the most detailed observations of neutral hydrogen across a wide range of redshifts, offering unprecedented insights into its cosmic evolution. The order-of-magnitude improvement in sensitivity introduces not only new opportunities for groundbreaking science but also challenges, necessitating the development of innovative analysis techniques to handle the complex datasets produced. 

This paper presents the results of the SKA Science Data Challenge 3a \footnote{\url{https://sdc3.skao.int/overview}} (SDC3a), which focuses on developing and testing foreground removal techniques for an EoR experiment using realistic synthetic datasets that emulate the capabilities of the SKA-Low telescope. A major challenge lies in the lack of a detailed model for Galactic synchrotron emission at spatial resolution and frequencies of the SKA-Low, making the removal of this emission particularly complex. Additionally, source confusion from previously unobserved extragalactic radio sources further complicates the task.

Foreground mitigation strategies generally fall into three categories: subtraction, avoidance, and suppression \citep[for an overview see][]{chapman19}. Foreground subtraction aims to remove contamination across all $k$-scales in Fourier space, increasing the range of scales available for analysis. However, inaccuracies in subtraction can introduce biases across all $k$-scales, potentially contaminating the signal. Foreground avoidance, in contrast, sidesteps contamination by restricting analysis to well-defined "clean" windows in cylindrical power spectrum ($k_\parallel$ vs. $k_\perp$), free from foreground influence. While this approach avoids subtraction-related biases, it limits the usable scales, potentially introducing its own biases in the averaged power spectrum. Finally, foreground suppression reduces the influence of contaminated $k$-scales by down-weighting them, mitigating the impact of residual foregrounds or subtraction errors. These strategies, individually or in combination, provide a framework for addressing the critical challenge of foreground mitigation in 21\,cm experiments.

In SDC3a, participants were tasked with recovering cylindrical power spectra of the redshifted 21\,cm signal, which is deeply buried beneath much stronger foreground emissions, using their own mitigation methods and strategies on realistic synthetic SKA-Low datasets. The primary goal was to accurately extract the redshifted 21\,cm signal while effectively mitigating foreground contamination and subtracting noise, ensuring that the results include robust error estimates. Participants were required to recover power spectra across various frequency segments within the simulation's frequency range, each corresponding to different redshifts. Where required by the teams, computing resources were offered by a list of computational facility partners to complete the challenge. Submissions were scored and ranked with a dedicated metric; additional metrics are also considered in this work for an in-depth analysis of the results. 

This paper is organized as follows: Sec.\ref{sec:datasim} describes the simulated data; %Sec. \ref{sec:hpcs} presents the computational facility partners; 
Sec. \ref{sec:teams} describes the analysis performed by the teams.  The results are assessed in Sec.\ref{sec:analysis} and final conclusions are drawn in Sec. \ref{sec:conclu}. 

\section{Simulated data}\label{sec:datasim}
A detailed description of the simulation pipeline used to produce the dataset is given in \cite{sdc3a1}. Here we only provide a short summary. The SDC3a simulation was undertaken with the {\tt OSKAR}\footnote{\url{https://ska-telescope.gitlab.io/sim/oskar/}} \citep{oskar} package, which makes use of a telescope model, which describes the instrument, and a sky model, listing all astronomical source components. There is provision to add several instrumental effects in the error model. 

The simulation corresponds to a 1000\,h observation. To reduce computational cost, a four hour duration observation was simulated, directed at (RA,Dec) = (00\,h, -30\,deg) using a 10 second integration time and a 100 kHz frequency sampling between 106 and 196\,MHz. The thermal noise level was finally rescaled to correspond to the target 1000\,h integration. This approach mimics accumulating 250 four hours observations to produce the final data product. The output of {\tt OSKAR} are visbilities.

%\item {\bf Images}. 
The {\tt WSClean} task \footnote{\url{https://gitlab.com/aroffringa/wsclean}}\citep{2014MNRAS.444..606O} was used to produce image cubes from the visibilities. Two different versions of an image and corresponding synthesised beam were produced: the first with \emph{natural} weighting and no deconvolution and the second with \emph{uniform} weighting followed by Gaussian tapering and a multi-scale deconvolution. 
%\end{itemize} 

\subsection{Telescope model}
The basic telescope model makes use of the SKA-Low configuration of 512 stations presented in
\cite{ska_low_conf}. The station layout is the so-called \emph{Vogel} layout, a one arm spiral configuration with a uniform areal density of antennas and a maximally diverse azimuthal sampling \citep{vogel}.

\subsection{Sky model} \label{sec:skymodel}
\begin{figure}
\includegraphics[width=8.5cm]{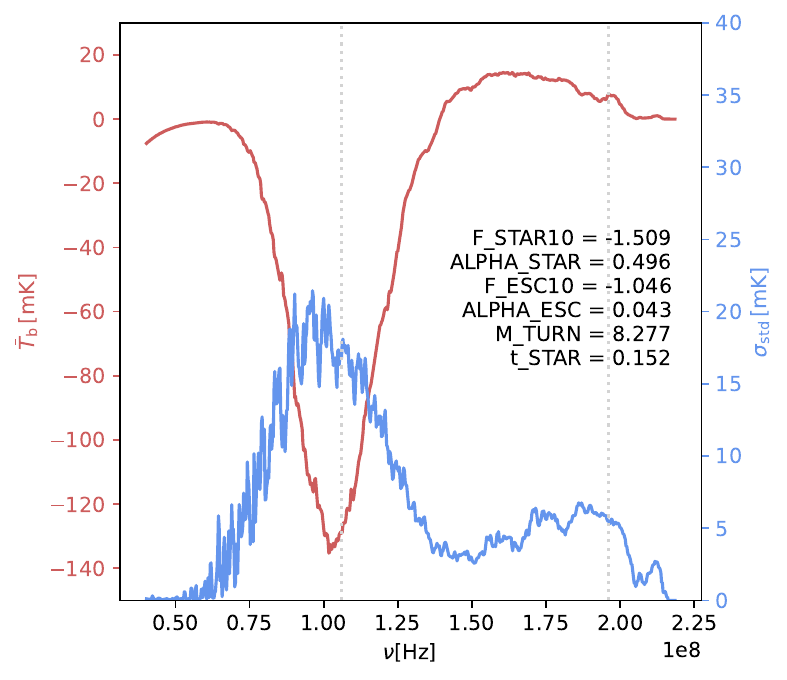}
\caption{{\tt 21cmFAST} parameters used for the generation of the EoR signal and corresponding mean (red) and standard deviation (blue) of the brightness temperature as a function of frequency.}
\label{fig:global}
\end{figure}

The sky components added to the simulations are the following:

\begin{itemize}
\item {\bf Out-of-field extragalactic sources}. Sources above 5\,Jy at 150\,MHz were added over the full 2$\pi$ steradians above the horizon at any time. Those are the so-called "A-Team" sources as well as sources from a composite GLEAM and LoBES catalogue \citep{2021PASA...38...57L}.

\item{\bf In-field extragalactic sources}. Within $8 \times 8$\,deg of the field centre, sources were added from GLEAM/LoBES above 100\,mJy at 150\,MHz, and from a mock T-RECS \footnote{\url{https://github.com/abonaldi/TRECS}}\citep{2023MNRAS.524..993B,2019MNRAS.482....2B} catalogue for $1\,\mu\rm{Jy}\leq I_{150\,\rm{MHz}}<100\,mJy$. In total we added over 15 million sources. 

\item {\bf Galactic emission}. We used a modified version of the GSM2016 sky model \citep{2017MNRAS.464.3486Z}, with a quadratic interpolation in log(frequency) to eliminate discontinuous derivatives within our band. 
Furthermore, spatial frequency content beyond the native resolution 
($\sim$1 degree at our frequencies) was added to the GSM2016 output 
using synthetic observations of low-frequency synchrotron emission from 
\cite{2022A&A...663A..37B}
Bracco et al. (2022). These synthetic observations were based on 
magneto-hydrodynamical (MHD) simulations of colliding super shells 
in the multiphase interstellar medium (ISM).

\item {\bf CD and EoR}. The cosmological signal has been produced with the {\tt 21cmFAST}\footnote{\url{https://21cmfast.readthedocs.io/en/latest/}} \citep{Murray2020,2011MNRAS.411..955M} code, for Planck18 cosmology $\Omega_m=0.30964$, $\Omega_\Lambda=0.69036$, $H_0=67.66$ \citep{2020A&A...641A...6P}. The {\tt 21cmFAST} reionizatoin parameters and the corresponding Brightness temperature are shown in Figure \ref{fig:global}. This scenario has been chosen as it yields a strong CD/EoR signal over a wide frequency range, therefore providing good potential for detection and measurement. The simulated EoR signal, in units of cubic Mpc, has been converted into observational quantities (deg, deg, MHz) by means to the {\tt tools21cm} library\footnote{\url{https://tools21cm.readthedocs.io/}} \citep{giri2020tools21cm}. 
\end{itemize}

\subsection{Error model}
In an effort to make the simulation more realistic, various instrumental errors were included. 

\begin{itemize}
\item{\bf Partial de-mixing}. The strong sources entering the field from the far sidelobes are normally modelled and removed with a so-called “de-mixing” process within a calibration and imaging pipeline. To simulate a partially successful de-mixing, the out-of-field sources were added with amplitudes attenuated by a factor of $10^{-3}$ \citep[see][for more details]{sdc3a1}.
\item{\bf Ionospheric effects}. We have used the {\tt ARatmospy}\footnote{\url{https://github.com/shrieks/ARatmospy}} \citep{2015OExpr..2333335S} code to construct an ionospheric model that was intended to represent moderately good observing conditions, characterised by a correlation scale, $r_0 = 7$\,km. The total electron counts and the corresponding phase modulation has been attenuated by the factor $10^{-2}$ to mimic the outcome of a successful direction dependent (DD) calibration. 

\item{\bf Gain calibration errors}. We have simulated the effect of direction independent (DI) gain calibration errors, by adding a noise term to the gain model, with standard deviations of 0.02 degrees in phase and 0.02\% in amplitude for each of the time and the frequency domains. The time domain represents residual broadband gain calibration errors and the frequency domain represents residual bandpass calibration errors.

\item{ \bf Thermal noise}. The nominal sensitivity per polarisation per unit bandwidth and unit time provided by {\tt OSKAR} was scaled finally to represent the average of two polarisations and a total observing time of 1000 hours, instead of the 1 polarization and 4\,h track that was simulated.
\end{itemize}

\subsection{Released data products}
The simulated dataset consisted of the following products:
\begin{itemize}
    \item Gridded visibilities, in the Measurement Set (MS) and UVFITS formats;
    \item Image cubes, with natural weighting and with uniform weighting and Gaussian tapering, and the associated Point-spread function (PSF) cubes;
    \item Time-averaged primary beam (PB) at all frequencies;
    \item Test dataset, consisting of an EoR + noise imaging simulated observation (with a different EoR model than what used for the challenge) and accompanying "true" power spectrum in the 166--181\,MHz frequency range, to help teams validate their power spectrum calculation pipelines. 
\end{itemize}

\subsection{Systematics introduced by the simulation pipeline} \label{sec:systematics}

Upon detailed inspection of the SDC3 dataset, an error in the simulation pipeline was uncovered, which meant that the sky components in the simulated data were not as intended. The issue involved the frequency behaviour of the EoR signal, the Galactic foregrounds and the faint extragalactic sources, which was consistently steeper than what modelled by $\Delta \alpha=-0.7$. 

Although the dataset has since been re-issued \citep[][presents the corrected simulation outputs and codes]{sdc3a1}, the problem was discovered only after the challenge completion. This work therefore makes use of the original SDC3a dataset and is scored against the corresponding "true" EoR. 

Although the issue described above affected the modelling of the components, it did not compromise the usefulness of the dataset for foreground removal.  One hypothesis that is often crucial to this part of the analysis is that the foreground components are smooth in frequency. The frequency smoothness of the components was not altered by the steepening, therefore there was no impact on foreground removal methods relying on this assumption.

While the simulated EoR signal strength was affected, there are only upper limits on this quantity, and the simulated signal is abundantly within the existing constraints. Similarly for Galactic synchrotron, which is by far the dominant diffuse foreground component at the frequencies considered, the committed error is within model uncertainties. A realistic Galactic synchrotron model should include both significant variation of the synchrotron spectral index across different regions of the sky \citep{2002A&A...387...82G, 2013A&A...553A..96D}, due to local variation in the magnetic field and  electrons density, and spectral steepening with frequency, due to electron energy losses \citep{2003A&A...410..847P}. However, the lack of data especially at low frequency provides limited constraints \citep[e.g.,][]{2016A&A...594A..10P,2016A&A...594A..25P}. %The spatial variations alone, which are not modelled in GSM2016, could account for difference up to $\Delta \alpha =0.7$ \citep{2002A&A...387...82G, 2013A&A...553A..96D}. 

Current constraints on the faint extragalactic source population are more abundant and stringent \citep[e.g.][]{LOTSS}. Source-by-source spectral index variations are significant, but there is no mechanism that would result in a consistent spectral steepening across an entire region of the sky. Therefore, the faint extragalactic source population in this work is not a good representation of the real sky. Amplitudes are consistent with the model at 106\,MHz but, due to the spectral steepening, are lower than what expected by about 65\%  at the highest frequency of 196\,MHz. 

Additionally, the issue described above affected the test dataset, although to a smaller extent due to the limited frequency coverage used for this product (up to 6\,\% at 181\,MHz). Teams that had used the test dataset to calibrate their final results were allowed to submit a correction after the error was discovered. Due to time constraints, only one team (HIMALAYA) submitted the correction.

\subsection{The challenge defined}
%The goal of the challenge is to measure the CD/EoR signal that is buried underneath the much stronger foreground emission. 
We asked the teams to recover six cylindrical (2D) power spectra $P(k_\parallel,k_\bot)$ of the EoR signal, clean from foregrounds and noise-subtracted, and the corresponding error bars. 
The 6 spectra were to be computed each over 15\,MHz intervals and without overlap, to cover the whole 90\,MHz frequency range of the simulation, and over the central 4 $\times$ 4 degrees out of the full 8 $\times$ 8 degrees field of view (FoV), to limit noise. The errors on the power spectrum were to be approximated as Gaussian and uncorrelated and provided as 1\,$\sigma$. The bins to be used to average the power spectrum in both $k_\parallel$ and $k_\bot$ were also provided.  

The computation of the power spectrum requires converting the observational units (deg, deg, MHz) back to comoving units (Mpc, Mpc, Mpc), using again the cosmological parameters. It is worth noting that, for this inverse conversion, we have given indications to adopt $H_0=100$ instead of $H_0=67.66$ as assumed in Sec.\ref{sec:datasim}. This is because power spectrum routines and libraries available in the literature can adopt either Mpc or Mpc$/h$ (with $h=H_0/100$) as units of comoving distance, which gives rise to inconsistent power spectrum binning and normalization. The choice of $h=1$ allows comparing power spectra with each other and with the ground truth, whichever convention was used for the comoving distance. %We note that this choice means that the spectra using this normalization do not correspond to the Cosmology used in the simulation, which adopts $h=0.6766$ instead. 
This is justifiable only in light of the scope of the challenge, which is foreground removal only and does not include inference of the reionization properties. 

\begin{figure*}
\includegraphics[width=16cm]{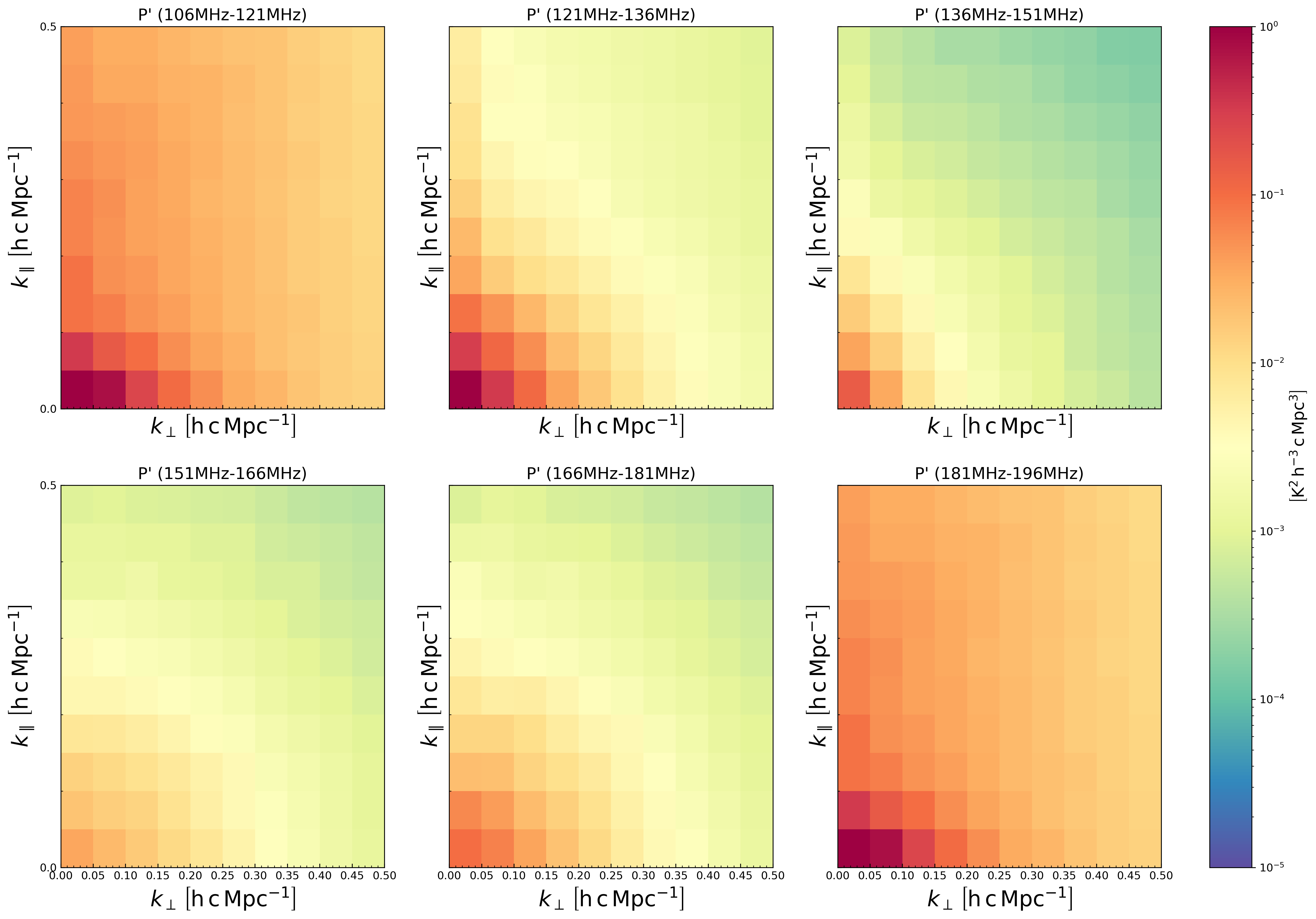}
\caption{Cylindrical power spectrum of the true noiseless EoR, $P'$ (computed with the $H_0=100$ convention)}
\label{fig:truth}
\end{figure*}

The 2D power spectra for the true noiseless EoR signal were computed with the {\tt tools21cm} library using the frequency and $k_\parallel, k_\bot$ binning adopted in this challenge and the $h=1$ convention.
The resulting true EoR power spectrum for all considered frequencies in shown in figure \ref{fig:truth}; this has been used for the computation of the challenge score and other performance metrics. 

\begin{figure*}
\includegraphics[width=16cm]{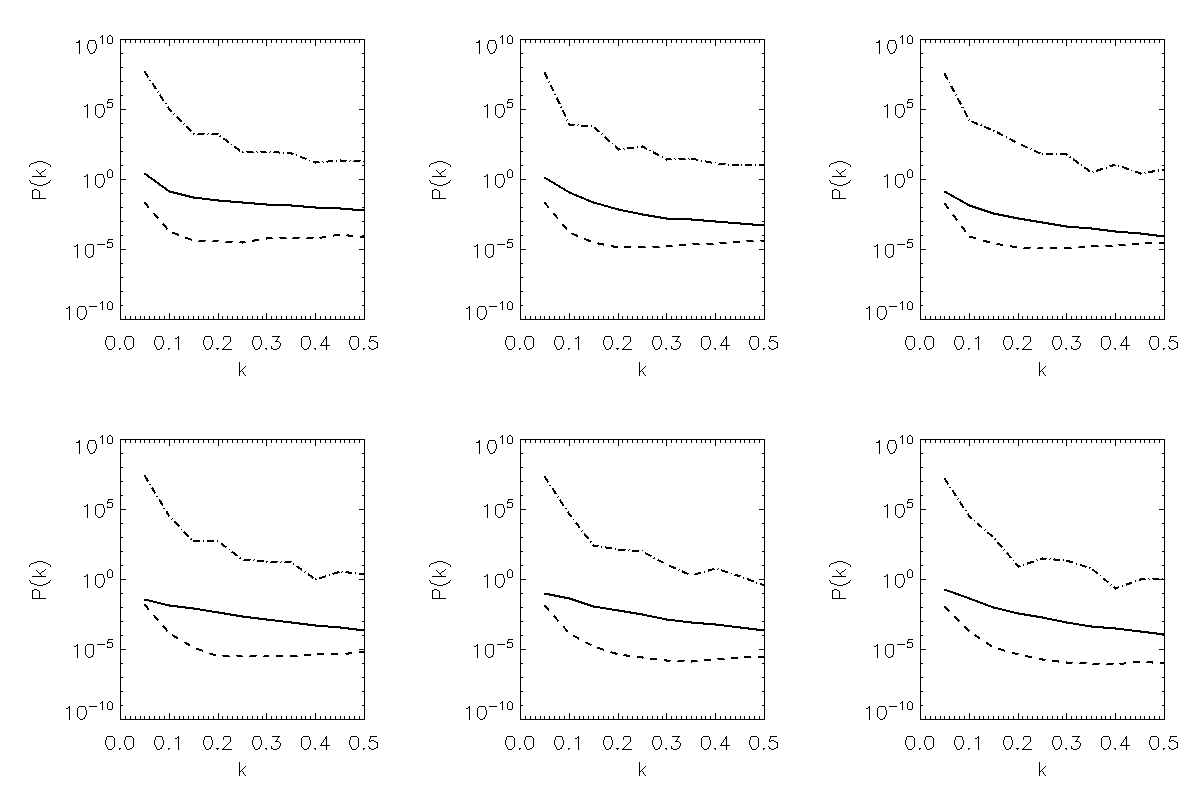}
\caption{Diagonal terms ($k_\parallel=k_\bot$ elements only) of the power spectra or the simulation (dot-dashed line), clean EoR (solid line) and instrumental noise (dashed line) for all the frequency channels (Top: 106--121\,MHz, 121-136\,MHz, 136-151\,MHz) from left to right. Bottom: 151--166\,MHz, 166--181\,MHz, 181--196\,MHz from left to right.}
\label{fig:data}
\end{figure*}

Figure \ref{fig:data} compares the EoR power spectrum (solid line) with that of the simulated data (dot-dashed line) and instrumental noise (dashed line). In order to compare them on the same plot, only the diagonal terms of the 2D power spectra ($k_\parallel=k_\bot$ elements) are shown, as representative of the others. 
This figure illustrates the challenge faced by the teams in reducing a foreground contamination many orders of magnitudes greater than the signal of interest. It also shows that, given the choice of a simulated 1000\,h exposure, the contamination due to noise is much more limited. 

\subsection{Computational facility partners} \label{sec:hpcs}
Due to the computational complexity of the challenge, an effort was made to allow all interested teams to participate, by offering dedicated computational resources to teams that requested them. This was done though a network of computational facility partners, that kindly made some resources available to SDC participants. The network included various computing facilities, having in general different hardware and software infrastructure. Globally, they were able to offer a diverse set of resources, including CPU and GPU computation, and different specifications regarding virtual and physical memory. 

No attempt was made to deliver a uniform resource allocation across different teams. Instead, teams specified their needs and were awarded the allocation that matched their needs more closely, subject to availability. Teams that worked on their own resources were able to download the data from one of the facility partners. Given the size of the visibility set (7.5\,TB for each of the MS and UVFITS formats), this was much more efficient than having a single download point. 

The list of facility partners in alphabetical order, and a brief description of the provision they offered, is below.

%\begin{itemize}
%\item Example. Resources were made available on the Servername HPC, broad specs. Any other relevant info. Acknowledgements. 

\subsubsection{ASTRON / SURF (Amsterdam, Netherlands)}
The ASTRON/SURF facility provided access to Spider\footnote{\url{https://doc.spider.surfsara.nl}}, a versatile high-throughput data-processing platform designed for processing large, structured data sets. Spider is a combined CPU and GPU system, located at the SURF data centre in Amsterdam, the Netherlands. Access to this system was supported by FuSE, the Fundamental Sciences e-Infrastructure, Dutch Research Council (NWO) Roadmap Project 184.035.004, and by SURF Innovation Funds.

%\subsubsection{AusSRC (Perth, Australia)}
\subsubsection{China SRC (Shanghai, China)} 
The China Square Kilometre Array (SKA) Regional Centre prototype (cnSRC) \citep{2019NatAs...3.1030A,2022SCPMA..6529501A} is a cutting-edge computing platform that employs a novel hybrid architecture.
This advanced setup integrated general-purpose Intel x86 CPUs, GPUs, and ARM processors, offering unparalleled flexibility and efficiency for data processing tasks.
The prototype comprises 35 Intel x86 CPU nodes with a total of 2240 cores (up to 128 cores per node), 12 ARM CPU nodes with 1152 cores (96 cores per node), and 4 GPU nodes equipped with 16 Nvidia V100 and 8 A100 GPUs,
collectively delivering approximately 800 TFlops of computing power. The CPU nodes contribute an additional 196 TFlops of computational capacity \footnote{\url{https://shaoska-user-guide.readthedocs.io/}}.
In terms of storage, cnSRC offers an approximately 9\,PB of distributed storage capacity, supported by a high-speed internal network enabling data exchange at rates of 100-200\,Gb/s between compute and storage nodes. Certain nodes are capable of accommodating up to 4\,TB of memory, with a maximum of 36\,GB per core, ensuring efficient processing of large-scale datasets.
Furthermore, cnSRC is connected to a 2\,Gbps transcontinental internet link, facilitating seamless international collaboration, and maintains a 200 Mbps connection with other SKA regional centre nodes to support global research initiatives.

This work used resources of the China SKA Regional Centre prototype funded by the Ministry of Science and Technology of the People's Republic of China and Chinese Academy of Sciences, National SKA Program of China (Grant No. 2022SKA0130103, 2022SKA0120102). Shaoguang Guo is supported by the international partnership program of the Chinese Academy of Sciences (Grant No. 018GJHZ2024025GC).

\subsubsection{Galicia Supercomputing Center/ CESGA (Santiago de Compostela, Spain)}
The Galician Supercomputing Center (CESGA) facility provided access to FinisTerrae-III \footnote{\url{https://cesga-docs.gitlab.io/ft3-user-guide/overview.html}} supercomputer, located in Santiago de Compostela, Spain. FinisTerrae-III is an HPC system integrating Intel x86 CPUs, AMD CPUs and Nvidia GPUs, with over 384 nodes, totalling more than 22,000 x86 cores and 157 GPUs with Infiniband interconnect and 6 PB of high performance storage Lustre Filesystem. 
For SKA SDC3a, CESGA allocated up to 100.000 core-hours and up to 2\,TB of permanent storage per team.
The supercomputer FinisTerrae III and its permanent data storage system have been funded by the NextGeneration EU 2021 Recovery, Transformation and Resilience Plan, ICT2021-006904, and also from the Pluriregional Operational Programme of Spain 2014-2020 of the European Regional Development Fund (ERDF), ICTS-2019-02-CESGA-3, and from the State Programme for the Promotion of Scientific and Technical Research of Excellence of the State Plan for Scientific and Technical Research and Innovation 2013-2016 State subprogramme for scientific and technical infrastructures and equipment of ERDF, CESG15-DE-3114.

\subsubsection{GENCI / IDRIS (Orsay, France)}
GENCI (\textit{Grand Equipement National de Calcul Intensif}) granted CPU and NVIDIA V100 GPU computing resources to two teams on the Jean Zay supercomputer hosted by IDRIS (Institute for Development and Resources in Intensive Scientific Computing).

IDRIS is the national supercomputer center of the CNRS (French National Center for Scientific Research) for High Performance Computing and Artificial Intelligence. It currently hosts Jean Zay, a supercomputer comprising 720 Intel CPU nodes and 843 GPU nodes equipped with V100, A100 or H100 NVIDIA GPUs for a total of 3704 GPUs. Following three successive extensions, the cumulated peak performance of Jean Zay reached 126 Pflop/s starting in July 2024.

\subsubsection{INAF (Italy)}
The Italian National Institute for Astrophysics (INAF), through its “Unità Scientifica Centrale 8” (USC8), provided computational support via the distributed HPC facility “Pleiadi.” This infrastructure consists of 192 nodes, totaling over 7,000 Intel Xeon E5-2697 V4 cores, interconnected via an Omni-Path HFI Silicon 100 Series network at 100 Gbit, with external connectivity at 1 Gb/s, and managed through the SLURM workload scheduler.  

For SKA SDC3a, INAF allocated up to 8 CPU nodes (128 GB RAM each) at the Institute of Radioastronomy in Bologna and up to 6 CPU nodes (128 GB RAM each, each with an available GPU) at the Astrophysical Observatory of Catania. Each team granted access to INAF resources had a dedicated 10 TB storage space for the duration of the challenge, ensuring seamless execution of workflows and analysis. Additionally, the challenge input data, including visibilities in both UVFITS and MS formats and image cubes, were made available in a read-only partition at both sites, facilitating efficient data access and processing.

\subsubsection {JPSRC (Tokyo, Japan)}
The Japanese SKA Regional Centre (JPSRC) prototype has provided computing resources for SKA SDC3a. JPSRC is an inter-university cooperation led by the SKA1 Japan Promotion Group (SKAJ) of the National Astronomical Observatory of Japan (NAOJ). JPSRC contains heterogeneous workstations with 152 CPU cores, 6 GPUs, 1 NEC SX-Aurora TSUBASA vector engine, as well as 1.86\,TB DRAM memory, 3.71\,TB M.2/SSD, and 338\,TB HDD storage as of the SDC3a. The workstations were federated by OpenStack as virtual machines, but JPSRC solely provided one of the workstations as a stand-alone machine to segment the servers into the one for SDC3a and the others for deploying SRCNet v0.1. 

The provided workstation was equipped with an AMD EPYC 7713P 64 core CPU, 2 x NVIDIA Quadro A6000 48GB, 1.024\,TB DRAM memory, and 162 TB HDD, and the network speed was typically 1\,Gbps. The operating system was Linux Ubuntu 22.04, and an environment for docker and singularity containerizations was supported. An SSH access was only permitted, and the primary data cube provided by SKAO for SDC3a was pre-stored by JPSRC at the local storage, to which a team was able to access.

\subsubsection {SPSRC/IAA-CSIC (Granada, Spain)}
The Spanish prototype of an SRC (a.k.a. espSRC), being developed at the Instituto de Astrofísica de Andalucía (IAA-CSIC) in  Spain,  is one of the 17 national initiatives contributing to the SRCNet development.  The espSRC OpenStack cloud gathers 240 CPUs cores and 2.5\,TB of memory across five compute hypervisors, plus 600+ TB of SSD usable storage capacity managed by Ceph. The servers are interconnected by a 100\,Gbps network and the cluster is connected to RedIRIS (the Spanish National Research Network ) with a 10\,Gbps link. 

The espSRC hosted two SDC3a teams which were provided with a virtual machine with 32 cores and 128\,GB memory. 100\,GB root disk using local SAS SSD and up to 5\,TB of additional block storage were assigned. The Operating System and the software in the virtual machine was fully customizable by the team through pseudo-sudo access. Accessibility to the machine was flexible, provided by ssh, remote desktop and a clientless remote desktop gateway.

The SPSRC acknowledges financial support from the State Agency for Research of the Spanish MCIU through the "Center of Excellence Severo Ochoa" award to the Instituto de Astrofísica de Andalucía (SEV-2017-0709) and from the grant CEX2021-001131-S funded by MCIU/AEI/ 10.13039/501100011033 \citep{SPSRC_cite}.

J. Sanchez and M. Parra acknowledge financial support from the grant CEX2021-001131-S funded by MCIU/AEI/ 10.13039/501100011033 and from the grant PID2021-123930OB-C21 funded by MCIU/AEI/ 10.13039/501100011033 and by ERDF/EU. J. Sanchez and M. Parra acknowledge financial support from the grant TED2021-130231B-I00 funded by MCIU/AEI/ 10.13039/501100011033 and by the European Union NextGenerationEU/PRTR.

\subsubsection {Swiss National Computing Center/ CSCS (Lugano, Switzerland)} 
The Swiss National Supercomputing Center (CSCS) is a key partner in deploying and operating the Swiss SRCNet node, contributing its expertise and computational resources to support scientific research. For this project, one SDC3a team was granted access to Piz Daint, CSCS’s former flagship supercomputer, which served as a testbed for Swiss SRCNet node science activities.  

At its peak, the now decommissioned Piz Daint was one of the most powerful supercomputers in Europe, comprising over 5,000 NVIDIA P100 GPGPUs. The team was allocated 15,000 node hours on this system, utilizing nodes equipped with one NVIDIA P100 GPU, 12 Intel Xeon CPU cores, and 65 GB of RAM. To support large-scale computations and data-intensive workflows, the project leveraged over 15 TB of long-term storage along with up to 10 PB of scratch space for high-speed temporary data access.  

This effort was made possible through the Performance Contract 2021-2024 in support to the activities of the Swiss CTAO Collaboration and the Swiss SKAO-Consortium funding with the Swiss State Secretariat for Education, Research and Innovation, SERI.

\subsubsection {UC-LCA (Coimbra, Portugal)}
The Laboratory for Advanced Computing of the University of Coimbra (UC-LCA) has provided access to its Navigator cluster, which includes CPU and GPU nodes. The resources included 20\,TB of shared storage space (including image cube data) and 100,000 core-hours for the team assigned to UC-LCA. The team could use up to 1032 cores of Intel Xeon E5-2697v2 or up 400 cores of Intel Xeon Gold 6148. Each node with these processors has 96 GB RAM and the interconnect is infiniband FDR/EDR. 
\subsubsection {UKSRC, IRIS-CAM (Cambridge, UK)}	
The UK SKA regional centre (UKSRC) resource at the University of Cambridge comprises a multi node HPC/GPU cluster running a Slurm batch scheduler. An OpenStack-hosted Platform-as-a-Service Azimuth applications portal is also available on request. Each SDC3a team was allocated 100,000 core hours, 1000 GPU hours and 20\,TB of storage.

Support comes from the Science and Technology Facilities Council (STFC) and the IRIS project.

\subsubsection {UKSRC, IRIS-MAN (Jodrell Bank Observatory, University of Manchester, UK)} 
Access was provided to the Galahad HPC cluster. This is a hybrid CPU/GPU cluster that includes large memory nodes that target wide field imaging and advanced image analysis. Support comes from the Science and Technology Facilities Council (STFC) and the IRIS project.

\section{SDC3a teams} \label{sec:teams}
This section presents the participating teams in alphabetical order, and describes the methodology used. 

\subsection{Akashganga (Arnab Chakraborty, Suman Chaterjee, Samir Choudhuri, Samit Pal, Narendra Nath Patra, Asif Elahi, Madhurima Choudhury, Anshuman Tripathi, Chandrashekhar Murmu, Srijita Pal, Santanu Das, Rajesh Mondal, Abinash Kumar Shaw, Rahul Shah, Gurmeet Singh, Soumadeep Saha, 
Utpal Garain)}
The team consists of observational cosmologists based in India, as a part of the SKA-India consortium. We employed a hybrid strategy to subtract the foreground and estimate the cosmological 21\,cm power spectrum.

We first made a high-resolution image of the visibility data using  {\tt WSClean} \footnote{\url{https://sourceforge.net/p/wsclean/wiki/Home/}} \citep{2014MNRAS.444..606O}. We made a large image of size $9.1^{\circ} \times 9.1^{\circ}$, using 2048 $\times$ 2048 pixels with a size of $16\arcsec$, using a uniform weighting. A 2nd order polynomial was fitted to the clean components to account for the frequency variation of the flux of the sources. The final model image was inverted and written to the MODEL\_DATA column. We used the UVSUB task in CASA \footnote{\url{https://casa.nrao.edu}} \citep{McMullin2007ASPC..376..127M} to subtract the continuum model from the visibility data. After continuum subtraction, we made a dirty image cube for each sub-band using the same parameters as provided in the SDC3a data description document. We made two image cubes using even and odd timestamps and cross-correlated them to estimate the power spectrum to avoid noise bias. 

For the residual foreground subtraction, we employed Gaussian Process Regression (GPR) \citet{Mertens18} and applied it to the point source subtracted residual data cube. GPR builds the model covariance of the different components of the data - foreground, 21\,cm signal, and noise. The hyperparameters of the covariance functions are optimized based on different spectral coherence of different signal components of the data. The foreground signal is then predicted from the model covariance of foreground and subtracted from the data cube. We applied GPR to both even and odd datacubes and then cross-correlated these datacubes and performed cylindrical averaging to estimate $P(k_\perp,k_\parallel)$. We corrected for signal loss by estimating the transfer function following the process outlined in \citet{Cunnington2023MNRAS.523.2453C}.

\subsection{Cantabrigians (Yuchen Liu, Oscar Sage David O'Hara, Eloy de Lera Acedo, Jeremy Coles, Samuel Alan Kossoff Leeney, Thomas Gessey-Jones, Dominic Anstey)}

Leveraging prior experience in data analysis for single-antenna 21\,cm cosmological experiments, we formed the Cantabrigians collaboration for the first time in this data challenge to develop a pipeline suitable for interferometric experiments with the SKA. Here we base the pipeline on a hybrid strategy that combines foreground removal and avoidance approaches. The pipeline takes the natural weighted images as input data and removes the foregrounds from the total sky intensity by point source subtraction and a Bayesian GPR model. The GPR approach our pipeline is based on is previously established in \citet{Mertens18,mertens20}. The underlying principle behind applying GPR to foreground removal is also explained in these works. The Bayesian framework within which the model is utilized is well described in \citet{Soares2022}. We further reduce the impact of astrophysical foregrounds by filtering the Fourier modes of the foregrounds from those of the 21\,cm signal, leaving an "EoR window'' in cylindrically-averaged 2D power spectra.

As our pre-calibration, bright point sources are removed from the SDC3a image data since we find the GPR models struggle with discrete radio sources scattered across the sky. The point source catalogue is a composite of the A-Team and MWA GLEAM and LoBES sources, provided in the data challenge. A sky model is created from this catalogue, applying the same flux filtering as described in Section~\ref{sec:datasim}. The model is then passed to the {\tt OSKAR} simulator\footnote{\url{https://ska-telescope.gitlab.io/sim/oskar/}} using an SKA-Low End-to-End Simulation Pipeline released in \citet{2024arXiv240204008O}. The settings for telescope layout, integration time, and frequency channels are also adjusted to maintain consistency with the SDC3a simulation. The measured visibilities generated from {\tt OSKAR} are gridded and inverse Fourier transformed to yield natural weighted images for the point sources by using {\tt WSClean}\footnote{\url{https://sourceforge.net/p/wsclean/wiki/Home/}}. The desourced images are obtained by subtracting the GLEAM and LoBES sources from the SDC3a image flux. 
Since the subtraction can be performed equivalently either in image or visibility space, we perform the subtraction in image space for reasons of efficiency.

For our GPR foreground subtraction, we use three kernel functions of the Mat\'ern class to represent the components in the observed sky signal: an exponential kernel for the 21\,cm signal, and two Radial Basis Function (RBF) kernels for both intrinsic and instrumentally corrupted foregrounds. The marginalized likelihood of the model is passed to a transdimensional nested sampler \texttt{PolyChord} \citep{2015MNRAS.450L..61H,2015MNRAS.453.4384H} to sample the posterior probability density and global evidence of the model. The hyperparameters with the peak posterior density are used in the model for signal separation. The global Bayesian evidence can also give a measure of how well the multi-kernel GPR model describes the observed data.

From the residual flux, the cylindrical power spectrum is calculated by using the {\tt ps\_eor} python code\footnote{\url{https://gitlab.com/flomertens/ps_eor}}. A tukey and a nuttall window function are applied to the data while transforming the images along their spatial axes and along the frequency axis respectively. The foreground-dominated regime draws a sharp boundary from the EoR window due to the frequency smoothness of radio frequency band foregrounds. This region is subsequently masked in the 2D power spectra based on the foreground avoidance strategy.

Here we discuss several potential improvements to the existing pipeline from three perspectives. Firstly, from an algorithmic aspect, the pipeline could incorporate a baseline-dependent foreground model for the separation of the 21\,cm signal. However, evaluating such a model for each baseline would incur significant computational expense. It is recommended to group the baselines by the length so that within each group the length scale variation is limited. Secondly, we have not used the CLEAN algorithm that is often employed to reduce the effects of dirty beam on the radio sources. Thirdly, uniformly-weighted images may be preferable over naturally-weighted ones in scenarios where the number of short baselines is significantly greater than the number of longer baselines, even though this approach comes at the cost of reduced sensitivity. Lastly, although we found gain instabilities and a phase screen to have negligible effects on the sources, corrections may be also needed in the pipeline, as they may result in source flux and position errors \citep{2018ApJ...867...15T,mertens20}. The full pipeline used in the SKA SDC3a can be found from our github page\footnote{\url{https://github.com/ycliu23/Cambridge-SKA-SDC3-Foregrounds}}, with a flowchart illustrating the aforementioned steps involved.

%%%%%%%%%%%%%%%%%%%%%%%%
\subsection{DOTSS-21 (F. Mertens, K. Chege, A. Offringa, B. Gehlot, S. Munshi, S. Ghosh, A. Acharya, B. Semelin, L. Koopmans, E. Allys, S. Brakenhoff, E. Ceccotti, D. Cornu, J.M. Delouis, H. Gan, C. Höfer, I. Hothi, F. Levrier, R. Meriot, M. Mevius, M.A. Miville-Dechenes)}

To address the SDC3a challenge our team employed methodologies grounded in the well-established practices of the LOFAR-EoR and NenuFAR Cosmic Dawn experiments, adapted to address the unique challenges of the SKA-Low simulated dataset. In essence this methodology consists of first detecting and subtracting compact and diffuse sources of foreground emission. We then use Maximum Likelihood GPR (ML-GPR) to model the residual foregrounds to separate them from the 21\,cm signal, and we finally generate the cylindrical power-spectra of the recovered 21\,cm signal.

\subsubsection{Compact Sources and Diffuse Emission}
Our strategy for subtracting compact sources and diffuse emission was executed iteratively. Initially, we subtracted the brightest sources from the raw visibilities. Based on the dataset description, this included A-Team sources and prominent sources from the GLEAM catalogue~\citep{2019PASA...36...47H,2021PASA...38...14F,2022PASA...39...35H}. Using {\tt OSKAR}~\citep{oskar}, we simulated visibilities for the six brightest A-Team sources (Taurus A, Cygnus A, Cassiopeia A, Fornax A, Pictor A, and Hercules A) alongside 210 point sources from the GLEAM catalogue, each exceeding 10\,Jy at 200\,MHz, adjusting their flux by a factor of $10^{-3}$ as outlined in the challenge documentation. This model of the "brightest" sources was then subtracted from the raw visibility data.
We proceeded to construct a sky model of "bright" compact sources within the primary beam's main lobe. Using {\tt WSClean}~\citep{2014MNRAS.444..606O}, we generated images of 2500 $\times$ 2500 pixels at a resolution of 15 arcseconds per pixel, imposing a minimum baseline restriction of 250 lambda to mitigate diffuse emission impact. The {\tt PyPBSF} tool was used to create an intrinsic sky model from these image cubes. The visibility model for these "bright" sources was predicted with {\tt OSKAR} and subtracted. We repeated this process on the residual data to model and subtracted "faint" point sources.
The final phase involved constructing a model for the Galactic diffuse emission using multi-scale cleaning techniques and a combination of Gaussian components for accurate representation~\citep{Offringa2017,Gehlot2022}. To ensure that the 21\,cm signal is not over-fit, we only used seven channels spanning the observation's full bandwidth, with steps of 15\,MHz. Employing again {\tt OSKAR}, we subtracted this model from the visibilities.

\subsubsection{Residual Foreground Removal}
The separation of the residual foregrounds from the 21\,cm signal was achieved through the application of the ML-GPR method, as described in~\cite{Mertens18,Mertens2024} and \cite{Acharya2023}.  This method separates the different components of the observed signal - foregrounds, the 21\,cm signal, and noise - by building a statistical model that exploits their distinct spectral-coherence signatures. We employ preexisting knowledge, either analytically derived or learned from simulations, to precisely separate these elements. In this application, the foreground emission was modeled using two components: one for the spectrally smooth intrinsic foregrounds and another for the 'mode-mixing' component (a consequence of the instrument’s chromatic response). Time difference image cubes provided a noise estimate. For the 21\,cm component, we used a kernel trained from {\tt 21cmFAST} simulations~\citep{2011MNRAS.411..955M,park2019inferring,Murray2020}, as described in~\cite{Mertens2024}.

\subsubsection{Power-Spectra Calculation}
The power spectra calculation employed the \texttt{pspipe}\footnote{\url{https://gitlab.com/flomertens/pspipe}} and {\tt ps\_eor}\footnote{\url{https://gitlab.com/flomertens/ps_eor}} 
packages, specifically developed for the analysis of 21\,cm observations within the LOFAR-EoR and NenuFAR Cosmic Dawn collaborations~\citep[e.g.][]{mertens20,Munshi2023}. Our team submitted three distinct results, each representing a different strategy in foreground management and signal extraction:

\begin{description}
  \item[\textbf{Avoidance strategy:}] Here ML-GPR was not run, and we instead favoured the filtering out of power spectra modes predominantly influenced by foregrounds.
  \item[\textbf{Removal strategy I:}] Here we focussed on the complete removal of foregrounds, applying ML-GPR to an image cube after the subtraction of only compact sources.
  \item[\textbf{Removal strategy II:}] An extension of Removal strategy I that included the subtraction of the modeled diffuse emission as well. This approach aimed to improve the reliability of ML-GPR by reducing the dynamic range of residual foregrounds, albeit introducing an element of risk.
\end{description}

Ultimately, the two removal strategies, which implemented a full foreground removal, scored higher than the avoidance strategy, underscoring the effectiveness of this approach.

%%%%%%%%%%%%%%%%%%%%%%%%%
\subsection{ERWA (Z. Zhu, H. Shan, Q. Zheng)}
Given the potential for recovering the EoR signal through methods such as Principal Component Analysis (PCA), polynomial fitting or Independent Component Analysis (ICA), especially in scenarios where observational and instrumental effects (in particular, the frequency-dependent one) are not considered (e.g., \citealt{Chapman12}, \citealt{wang13}, \citealt{He24}), we propose the following methodology. Beginning with the "dirty" image cube, we employ a U-Net neural network for denoising and deconvolution, with the goal of revealing the underlying sky map. We then apply polynomial fitting to derive the EoR map and calculate its 2D power spectrum. 

\subsubsection{Beam deconvolution and data denoising}
The most critical part of our approach lies in the extraction of a clean sky map, free from instrumental and observational distortions, from the dirty map.
To achieve this, we propose the construction of a U-Net network. The architecture of our U-Net network is designed to be straightforward yet effective, comprising 6 convolutional layers, 3 max-pooling layers, 6 transpose convolutional layers, 3 up-sampling layers, and a final transpose convolutional layer. The network inputs the dirty map with instrumental and observational noise and outputs the corresponding clean sky map.

For the training set, we construct the sky map following \citet{li19}, incorporating components such as Galactic synchrotron emission, Galactic free-free emission, extragalactic point sources, radio halos, and the EoR signal. Utilizing the {\tt OSKAR} software package, we simulate the SKA-Low antenna's response and layout, incorporating thermal noise, antenna gain and phase errors, and ionospheric effects. The dirty map is generated using the {\tt WSClean} package, with simulation parameters such as FoV, pixel size, and frequency resolution aligned with official specifications.

In the pre-training phase, bright point sources are masked. The dirty maps are then standardized across frequencies and segmented into 128x128 pixel squares. Training then proceeds by minimizing the pixel-wise L2 norm between the network-predicted sky map and the true sky map.

\subsubsection{Foreground removal and Power-spectrum calculation}
We apply the U-Net network to the SDC3a natural weighting image cube to obtain the clean sky map. Subsequently, a pixel-by-pixel fifth-order polynomial fitting is performed on the sky map for each frequency range for the foreground removal.

It is important to note that the current performance of our pipeline is not without flaws. Some point source residues need masking, and in certain map areas, there is an absence of signal post-foreground removal. Thus, we mask regions with no signal (value $= 0$) and those with clear point source residue (value $> 0.5$ K). Following this, we directly calculate the 2D power spectrum for each frequency range and submit these results. However, it must be acknowledged that we ought to have computed the pseudo-power spectrum using this mask. Given that our final results are several times lower than expected across most bins, correcting this issue is anticipated to improve our outcomes.

\subsubsection{Future Plans}
Looking ahead, we aim to refine our methodology in three key areas. Firstly, we plan to enhance the network architecture to incorporate the PSF during training. Secondly, we intend to improve pre-training strategies for more effectively eliminating bright point sources or strong Galactic foregrounds, along with adopting more robust measures for evaluating training outcomes beyond the L2 norm. Lastly, we aim to advance our training set to encompass more realistic effects, thereby optimizing the training process.

The ERWA team acknowledges funding from the National Natural Science Foundation of China (No. 12203085) and the National SKA Project of China (No. 2020SKA0110100).
%%%%%%%%%%%%%%%%%%%%%%%%%%%%%
\subsection{Foregrounds-FRIENDS (M. Ruiz-Granda, I. Labadie-García, M. Cruz, M. Aparicio, A. Fernandez, M. López-Caniego, V. Salvador,  L. Darriba, J. Moldón, J. Garrido, D. Herranz, M.-L. Gendron-Marsolais, M. Remazeilles, J. A. Rubiño-Martin, S. Sánchez, L. Verdes-Montenegro, P. Vielva, T. Wiegert)}

The Foregrounds-FRIENDS team implemented a workflow \citep{Foregrounds-FRIENDS} based on PCA to perform the foreground removal of the data. The workflow, which is
publicly available in GitHub\footnote{\url{https://github.com/espsrc/FOREGROUNDS-FRIENDS}}, is coded mainly in {\tt Python} and can be installed with all the software
dependencies using {\tt conda} \citep{anaconda}. With this methodology, each step
can be developed, tested and executed independently from the others, facilitating modularisation and reproducibility of the workflow.

Our team is composed by a mixture of experts in Cosmic Microwave Background (CMB) science and in radio astronomy. To address this challenge we tried to combine methods developed within the framework of CMB research into this emerging field as it is Intensity Mapping. Our methodology can be decomposed in three steps: point source detection, foreground removal algorithms and power spectrum estimation.

\subsubsection{Point source detection}
Our initial idea was to translate the expertise of part of the team in point source detection and mask creation for CMB experiments like \textit{Planck} or QUIJOTE \citep[see e.g.,][]{quijote_mfi_radiosources} to this challenge. For that purpose point source catalogues were created to use them in the foreground removal step.

The point source detection procedure is a two-step process. First, we created lists of candidates for every frequency in the data cube using a well-established source detection software, {\tt SExtractor} \citep{Bertin96}. Second, we used convolution neural networks to assess the reliability for each candidate. The neural network structure and detection thresholds were adjusted to accommodate the unique conditions of the challenge, and a Mexican Hat wavelet kernel was used to remove the significant diffuse emission present in the images to improve the reliability assessment. 

This image segmentation problem aims at creating a cube of masks tagging pixels corresponding to point sources with a 1 and background pixels with a 0. To tackle this problem, a convolutional neural network with an autoencoder architecture was implemented, divided into two main components: the encoder and the decoder.

\textbf{Autoencoder Architecture:}
The encoder consists of three main convolutional layers. The first convolutional layer receives an image with a specific number of channels (\textit{input\_channels}), applies a convolution with a $3 \times 3$ kernel, generating an output with \textit{num\_filters1} channels, while padding ensures the spatial dimensions are maintained. This process is repeated in the second and third layers with \textit{num\_filters2} and \textit{out\_channels3} channels, respectively. After each convolution, the ReLU activation function is applied, followed by instance normalization and a MaxPool2d layer, halving the spatial dimensions. Finally, a Dropout layer is included to prevent overfitting.

The decoder is responsible for reconstructing the original image from the encoded representation, using a reverse process to that of the encoder with transposed convolution layers. The final layer applies the sigmoid activation function to ensure that the output values are between 0 and 1, indicating the probability of a point source's presence.

\textbf{Training and Optimization:}
The network was trained for 65 epochs using a set of 500 images, split into 80\,\% for training and 20\,\% for testing. Adam optimizer was used with an initial learning rate of $0.001$.

This methodology allowed for effective detection of point sources, facilitating foreground removal and enhancing the exploration capabilities of the background radiation.
As a result of this process, two additional products were generated: a single point source catalogue consolidating data from all frequencies, and a modified data cube where all detected point sources have been masked, facilitating the subsequent component separation of the Galactic diffuse emission and the cosmological signal.

The point source mask was not used in either of the two foreground removal algorithms explored. However, we consider it important to include it in this paper as it was a significant contribution from the team.

\subsubsection{Foreground removal algorithms}
Two different foreground removal algorithms were explored aiming to use both available data sets: cubes and visibilities.

The first one is based on PCA applied to the data cubes \citep{PCA_cleaning}. PCA is based on the eigenvalue and eigenvector decomposition of the frequency-frequency covariance matrix estimated using the data cubes. We kept the modes with the highest variance and subtracted them from the original cube. These modes contain most of the foregrounds, so by removing them we are ideally left with the 21\,cm signal. This analysis only uses frequency information and their correlation, not the spatial information of the data cube.

We applied a four component PCA to the data cube, which seems to be optimal from our simulations. The original data cube is 8 $\times$ 8 degrees FoV, but in order to limit the noise, only the central 4 $\times$ 4 degrees are used, as required by the power spectrum estimation.

The second approach is based on Polynomial fitting of the complex visibilities, following the procedure proposed in \cite{10.1093/mnras/stw2494}. We consider the total visibilities $\mathcal{V}(\mathbf{U}, \nu )$, and separately fit the real and imaginary parts as a function of the frequency $\nu$ for each Fourier space point $\mathbf{U}$. 
The fit is performed using a third-order polynomial in logarithmic space. The fitted model is then removed from the total visibilities to get the residual visibilities. The real-space maps are obtained by performing the inverse Fourier Transform of the residual visibilities at each frequency channel.

After comparing the performance of the two methods, we decided to use  the PCA approach for this challenge.

\subsubsection{Power spectrum estimation}
The cylindrical power spectrum has been estimated with the Python package {\tt ps\_eor}\footnote{\url{https://gitlab.com/flomertens/ps_eor}}. In our workflow, images from previous steps are converted to brightness temperature and the PSF is deconvolved. Appropriate bins that conform to the required submission format are determined to calculate the power spectrum, correcting for the primary beam. Noise-subtraction techniques have not been applied.

\subsubsection{Workflow and reproducibility}

The Foreground-FRIENDS workflow aims to provide a modular and flexible implementation. Each task in the workflow (except the point source detection) is implemented either as a Python script, a Jupyter Notebook or a Bash script. A generic and reusable python script launches all the tasks sequentially. The selection and order of the tasks can easily be modified through a configuration YAML file. Each task, or step, is stored individually in a separate directory containing relevant documentation explaining the purpose and methodology followed, algorithms used and parameters of each task.  For each task, we divided the contents in python scripts with the relevant functions, parameters files independent from the code, and notebooks to execute the functions and visualize the results, when possible. Therefore, we have aimed to isolate the algorithms, parameters and execution, thus facilitating reusability.

Software provenance is fixed using {\tt conda} through an explicit {\tt conda}environment. The repository contains license and citation information, as well as a README file with installation instructions, execution instructions, a description of implemented steps and a diagram of the file structure. For persistence and findability, the workflow is stored in the long-term general archive Zenodo.

The Foregrounds-FRIENDS acknowledges the Spanish Prototype of an SRC (SPSRC) service and support funded by the Spanish Ministry of Science, Innovation and Universities, by the Regional Government of Andalusia, by the European Regional Development Funds and by the European Union NextGenerationEU/PRTR. %The SPSRC acknowledges financial support from the State Agency for Research of the Spanish MCIU through the "Center of Excellence Severo Ochoa" award to the Instituto de Astrofísica de Andalucía (SEV-2017-0709) and from the grant CEX2021-001131-S funded by MCIU/AEI/ 10.13039/501100011033 \citep{SPSRC_cite}.

LVM, SS, JG, JM, TW, MGM, LD and IL acknowledge financial support from the grant CEX2021-001131-S funded by MCIU/AEI/ 10.13039/501100011033 and from the grant PID2021-123930OB-C21 funded by MCIU/AEI/ 10.13039/501100011033 and by ERDF/EU. LVM, JG, TW and MGM acknowledge financial support from the coordination of the participation in SKA-SPAIN, funded by the Ministry of Science, Innovation and Universities (MCIU). LVM, SS, JG, LD and IL acknowledge financial support from the grant TED2021-130231B-I00 funded by MCIU/AEI/ 10.13039/501100011033 and by the European Union NextGenerationEU/PRTR. IL acknowledges financial support from the grant PRE2021-100660 funded by MCIU/AEI /10.13039/501100011033 and by ESF+. JM acknowledges financial support from grant PID2023-147883NB-C21, funded by MCIU/AEI/ 10.13039/501100011033 and by ERDF/EU.

MRG, DH, MC, MR and PV have been supported by MCIN/AEI/10.13039/501100011033, project refs. PID2019-110610RB-C21PID2022-139223OB-C21 (funded also by European Union NextGenerationEU/PRTR), and from Universidad de Cantabria and Consejería de Educación, Formación Profesional y Universidades del Gobierno de Cantabria, via the "Actividad estructural para el desarrollo de la investigación del Instituto de Física de Cantabria". MRG acknowledges financial support from the Formación del Profesorado Universitario program of the Spanish Ministerio de Ciencia, Innovación y Universidades. MRG and DH thank the Spanish Agencia Estatal de Investigación (AEI, MICIU) for the financial support provided under the project with reference PID2022-140670NA-I00.

%%%%%%%%%%%%%%%%%%%%
\subsection{HAMSTER (Zhaoting Chen, Steven Cunnington, Aishrila Mazumder, Amadeus Wild, Laura Wolz)}
We develop a minimalist yet innovative approach to explore alternative methods for foreground mitigation without cleaning the foregrounds in the image cube. We aim to sufficiently mitigate wide-field effects in the visibility data, and perform direct foreground avoidance for measuring the power spectrum \citep{Morales12}. Therefore, our entire data analysis is performed in delay space without using the images at all. The analysis can be divided into two steps, which are flagging in delay space and coherent averaging for power spectrum estimation. We describe the method below.

\subsubsection{Flagging in Delay Space}
The visibility data can be Fourier transformed along the frequency direction into delay space. The foreground emission mostly resides within the wedge defined by the primary beam field-of-view \citep{2014PhRvD..90b3018L}. However, wide-field effects such as bright sources in beam sidelobes leak into higher delay and prohibits measurements of 21\,cm signal (e.g. \citealt{2016ApJ...819....8P}). We therefore identify and exclude the contamination at high delay first.

\begin{figure}[h]
    \centering
    \includegraphics[width=0.5\textwidth]{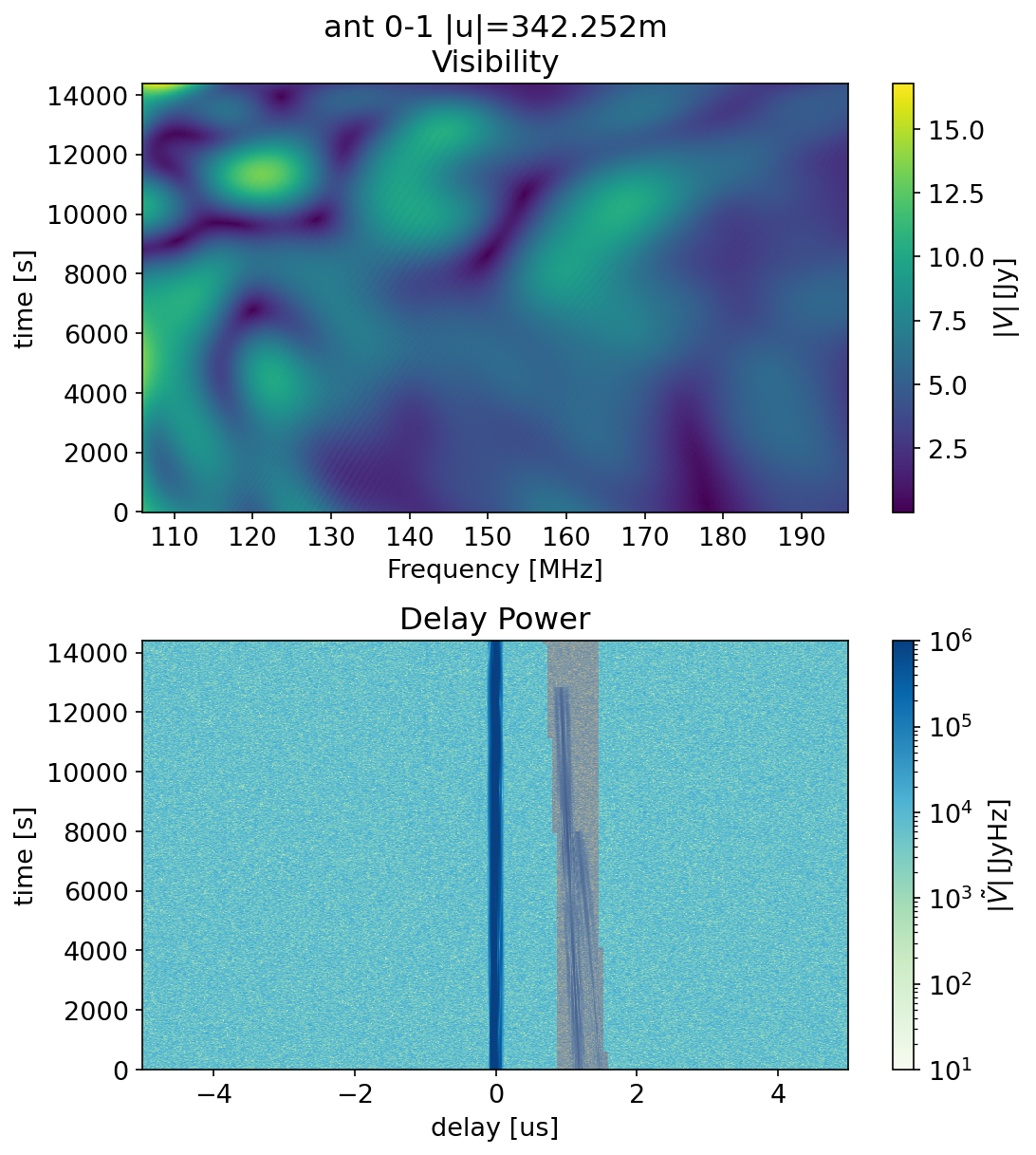}
    \caption{Illustration of the flagging method in delay space from the HAMSTER team. The top panel shows the amplitude of the visibility data for antenna pair 0 and 1. The bottom panel shows the amplitude of the delay-transformed visibility data for the same pair. Two visible tripes of excess power can be seen around the delay $\eta = 1\,\mu$s. The shaded region shows the delay modes that are excluded from the power spectrum estimation.}
    \label{fig:flag_illus_hamster}
\end{figure}

For each antenna pair, the visibility data is Fourier transformed. Delay power spectrum is then calculated as illustrated in Fig. \ref{fig:flag_illus_hamster}. Excluding the region around delay $\eta =0$ where foregrounds reside, the power spectrum in the rest of the time-delay space is used to perform iterative thresholding. The thresholding excludes 5-sigma of the sampling variance iteratively to find the average amplitude of the delay power, which is thermal noise dominated. The excluded delay modes, as shown in Fig. \ref{fig:flag_illus_hamster}, have clear, stripe-like structure that is continuous in time and delay. Based on the position of excluded modes in the time-delay plane, we perform a morphological closing of the flags to identify the stripes in each antenna pair. The stripes are then extended along the direction of time shown as the shaded region in Fig. \ref{fig:flag_illus_hamster}.

\subsubsection{Power Spectrum Estimation}
The flagged, delay-transformed visibility data are averaged into $u$-$v$ grids for power spectrum estimation. The visibilities are first averaged every 5 channels to match the maximum $k_\parallel$ required. A Blackman-Harris taper is then applied to perform the delay transform. The visibilities are then averaged into the grids with flagged delay modes excluded. We choose the grid length to be $3\lambda$. The dataset is divided into two sub-set for cross-correlation to remove the noise floor (see e.g. \citealt{2022ApJ...925..221A,2023arXiv230111943P}). The gridding is performed for the whole band for quality assessment, and then for each of the sub-bands required by the data challenge. The flags in delay space are downsampled to match the delay resolution of frequency sub-bands.

\begin{figure}[h]
    \centering
    \includegraphics[width=0.5\textwidth]{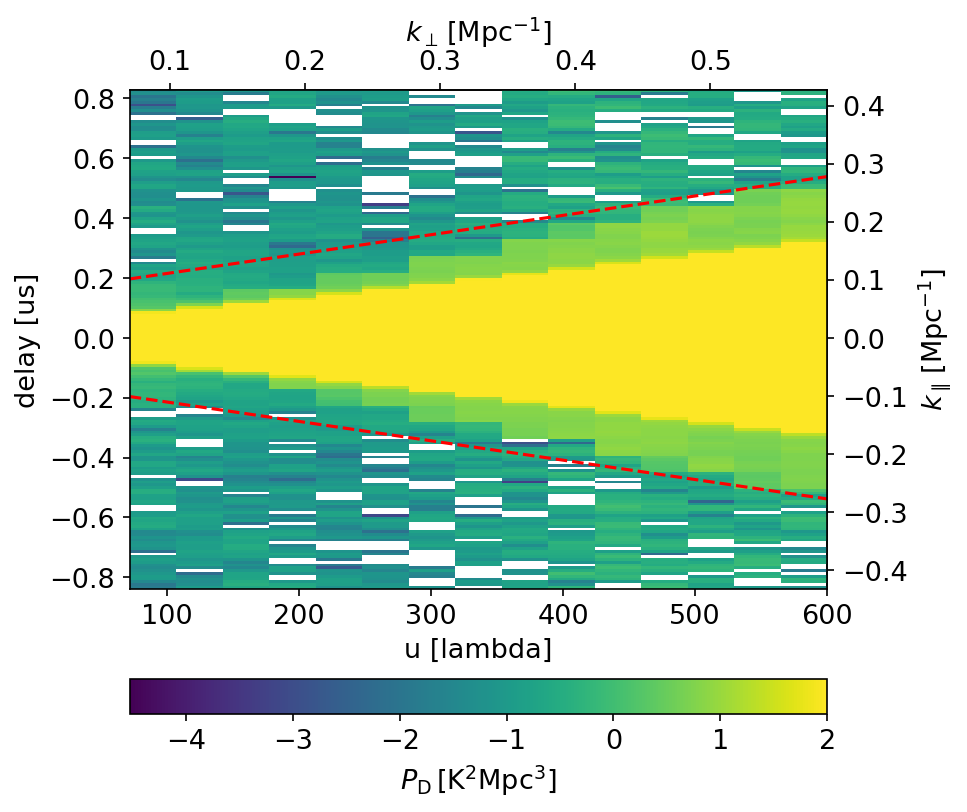}
    \caption{The estimated cylindrical power spectrum of the whole band from the hamster team. The red dashed line shows the region where foregrounds dominate.}
    \label{fig:ps_illus_hamster}
\end{figure}

The 3D power spectrum is then calculated as the square of the modulus of the delay-transformed gridded visibilities. Conversion to temperature units is performed with volume renormalization based on the power-squared beam area \citep{2014ApJ...788..106P}. The resulting power spectrum is shown in Fig.\ref{fig:ps_illus_hamster}. For the whole band, we find that the power spectrum is at $\sim 10^{-3}-10^{-2}\,{{\rm K^2 Mpc^3}}$, suggesting that signal level is being reached. However, the cylindrical power spectrum shows residual contamination exists when we perform the same estimation for each of the 6 sub-bands, when the measured power at $\sim 0.1\,{{\rm K^2 Mpc^3}}$. We postulate that this is due to the frequency taper not being able to suppress frequency-dependant gain errors injected at sub-band level.

\subsubsection{Outlook}
We plan to further develop on this method by adopting Bayesian statistics in the flagging of contamination in delay, coherently average visibilities in delay space, and power spectrum estimation. Specifically, we want to use the initial flagging and averaging to calculate angular scale-dependent data covariance. The covariance can then be used to inpaint the flagged delay modes and to repeat the averaging and power spectrum estimation. This can be performed iteratively so that the data covariance for inpainting and the subsequent power spectrum estimation converge.

Foreground removal can be additionally added to the data analysis. We plan to test frequency covariance-based cleaning such as PCA and GPR at different stages of the data analysis. This may further be implemented iteratively, where 21\,cm signal covariance and foreground covariance can be separately estimated.

%%%%%%%%%%%%%%%%%%%%%%%%%%%%%%%
\subsection{Hausos (Ming Jiang, Zhenzhen You, Jean-Luc Starck, Huanyuan Shan, Feng Shi)}
The team was composed of data scientists and cosmologists based in both China and France. With the experience in sparsity-based source separation methodology, we primarily intended to further extend the previously published DecGMCA~\citep{jiang2017joint}, starting from visibility data, to the SDC3a data challenge. Due to the various gridding strategies and large computation costs, we finally switched to a deep learning-based method. Since the classical convolutional neural network (CNN) is limited by its receptive field, we employed a transformer-based neural network based on SETR presented by \cite{zheng2021rethinking}. SETR is a model that combines transformer and CNN to achieve semantic segmentation tasks. With the transformer as the encoder, global contextual features are modeled in each encoder layer, which can effectively capture long-distance features. The decoder is a CNN-based model to upsample the advanced features extracted by the encoder to the original input resolution for pixel-level prediction.

\textbf{Training}
We used the EoR simulation from Shanghai Astronomical Observatory, established upon the {\tt 21cmFAST} simulations~\citep{2011MNRAS.411..955M,Murray2020} and {\tt OSKAR} using the same observation parameters. Since the final score was evaluated on the 6 frequency bins, each containing 150 frequency channels (except for the last bin containing 151 frequency channels), the training data were made to 6 bins as well. Thus, 6 independent networks for different bins need to be trained. In the training stage, the dirty maps and the ground-truth EoR maps were cut into patches of the size 256 $\times$ 256 pixels, and the network was trained with 100 epochs from dirty patches to the corresponding ground-truth EoR patches. The optimizer was Adam, and the learning rate was set to 1e-5. The binary cross entropy was chosen as the loss function.

%The simulation involves a large-scale box of 1.6 Gpc with 1024 grid cells, spanning a redshift range from 86.0 down to 5.0. Consistent with project recommendations, we adopt the "faint galaxy" lightcone for our analysis. To ensure compatibility with subsequent foreground maps, we extract image slices at specific redshifts and then carry out a precise tiling and scaling process. This procedure allows us to maintain a consistent image size across our data processing pipeline.We utilize the Fg21Sim software \footnote{https://github.com/liweitianux/fg21sim} (\citet{}) to generate four critical foreground components for our analysis: Galactic synchrotron, Galactic free-free, extragalactic point sources, and radio halo. We invite readers to delve into the specifics of Fg21Sim in \citet{}; here, we provide a brief overview of its application within our research.
\textbf{Prediction}
Given a frequency bin, the released natural weighting images, cut into $256 \times 256$ patches, were used as the input of the trained network, and the EoR patches were predicted for the frequency bin. Then the patches were concatenated to form the EoR images. Repeatedly for the 6 frequency bins, we finally obtained the EoR image cubes.
During the data challenge, we observed that the network did not perform well on the negative values due to the activation functions and the amplitude of the estimated power spectrum being biased. One reason may lie in the PSF not being modeled well enough, causing deconvolution problems. Thus, our future work will focus on incorporating the PSF modeling in the network.

MJ acknowledges funding support from the National Natural Science Foundation of China (Program No. 12203038) and the Guangdong Basic and Applied Basic Research Foundation (Program No. 2021A1515110057).
%%%%%%%%%%%%%%%%%%%%%%%%%%%%%%%%%
\subsection{HIMALAYA (Le Zhang)}
The team name HIMALAYA connotes that measuring the EoR 21\,cm signal is as challenging in astronomy as climbing Mount Everest. To estimate the EoR power spectra from the released SDC3a data, I proposed a new foreground-cleaning algorithm called the ``reconvolution'' method. This method involves three key steps: 1) reconvolution preprocessing of dirty maps (natural weighting) to recover the smoothness along frequency; 2) applying the PCA technique to the reconvolved dirty image; and 3) using the power spectrum correction to adjust the amplitude of the measured HI power spectrum at different $k$ bins. I will summarize them as follows.

\subsubsection{Reconvolution}
The aim of the reconvolution preprocessing is to greatly suppress the sidelobe of the PSF and reduce the so-called mode-mixing effect in interferometric observations. Without any instrumental effects, the foreground signal exhibits a very high degree of smoothness along frequency, which makes it easy to separate from the oscillating 21\,cm signal. Traditional techniques such as PCA or polynomial fitting can reliably recover the 21\,cm signal. However, due to incomplete $uv$-coverage and the mode-mixing effect from long baselines, the dirty image becomes complex and exhibits hard-to-model frequency dependence. Consequently, the measured foreground presents highly non-smooth features. The usual deconvolution methods, such as CLEAN, do not guarantee the smoothness of the sky map, potentially leading to inaccuracies in separating the 21\,cm signal. Moreover, due to numerical effects, any deconvolution attempt aimed at dividing the PSF over the Fourier modes results in unavoidable artifacts that prevent the 21\,cm signal recovery.

Considering the required $k$ range of about  $k< 0.5~h{\rm Mpc}^{-1}$ for SDC3a, detailed measurements on small scales are not necessary in principle. Therefore, a natural way is to convolve an appropriate beam kernel, $w(r)$,  for each frequency of the image. This process significantly suppresses the measured visibilities from the long baselines, while maintaining a spatial resolution sufficient to capture the Fourier modes within the desired $k$ range. In other words, following the convolution theorem, one can design a weighting scheme for the visibilities in the Fourier domain, aiming to suppress the high-$k$ Fourier modes while keeping the noise from being significantly amplified. From the tests, we find the easiest choice of the beam kernel is the SDC3a PSF.  This choice may not be optimal, and other options, such as using a Gaussian beam kernel, are also possible. Further optimization to improve the signal-to-noise ratio, more effective removal of foreground contaminants, etc., need to be further investigated through simulation. %However, it did work for the released data.

For SDC3a, there are 6 frequency bins, each containing 151 frequency channels. For each frequency bin, by convolving the PSF of the first channel with the 151 dirty maps, we found that the resulting reconvolved dirty image became highly smooth along frequency. Although the small-scale structures of the images are suppressed, the frequency smoothness is recovered, which is essential for the foreground removal. 

\subsubsection{PCA}
In the absence of instrumental effects, the eigenvalues of the frequency-frequency covariance of the sky image are expected to exponentially decrease. Typically, removing the first three eigenmodes by the PCA technique can effectively subtract most of the foreground contributions, leaving only a small foreground residual compared to the 21\,cm signal level. However, the eigenvalues of the raw dirty images decrease slowly, making the PCA technique ineffective. Even after subtracting approximately 120 modes in a SDC3a dirty map (given that we only have 151 modes in each frequency bin), we still have a difficulty in obtaining a residual image with a signal amplitude comparable to that of the 21\,cm signal. 

Surprisingly, after applying the reconvolution preprocessing, the distribution of eigenvalues becomes similar to that of the instrumental-free case. The total foreground contamination can be effectively cleaned by discarding the first 20-30 eigenmodes of the reconvolved dirty image, and the residual image pattern could reflect the correct cosmological HI pattern. After PCA, the 2D cylindrical power spectrum of the resulting image, $\tilde{P}(k_\perp,k_\parallel)$, can be estimated by averaging over the corresponding Fourier modes falling into the a given $(k_\perp$,$k_\parallel)$ shell. We used our own code to do this estimation. 

\subsubsection{Power spectrum correction}
There are four major factors that can alter the amplitude of the estimated power spectrum: the primary beam effect, PSF of the synthesized beam, reconvolution preprocessing, and 21\,cm signal loss due to the PCA subtraction. In our algorithm, we do not correct for these effects at the map level, but at the power spectrum level, by defining a transfer function $T(k_\perp,k_\parallel) = \left<P(k_\perp,k_\parallel)/\tilde{P}(k_\perp,k_\parallel)\right>$, where $P(k_\perp,k_\parallel)$ is our final submission. The noise bias is neglected in our analysis as it appears to be small compared to the 21\,cm signal. Then, the final reconstructed HI power spectrum for SDC3a is calculated using $P(k_\perp,k_\parallel) = T(k_\perp, k_\parallel) \tilde{P} (k_\perp, k_\parallel)$. 
The estimation of $T(k)$ is derived from the average value of the power-spectrum ratios obtained from mock data, produced as described below. In addition, to correctly estimate the error bar of $P$, one can also use the mock-derived standard deviation of $T(k_\perp, k_\parallel)$, $\Delta T(k_\perp, k_\parallel)$, to propagate all statistical and systematic uncertainties i.e., $\Delta P(k_\perp, k_\parallel) =\Delta T (k_\perp, k_\parallel) \tilde{P}(k_\perp, k_\parallel)$.

The transfer function does not explicitly rely on the precise shape of the input power spectrum (at least to good approximation), as it only depends on the ratio of the input to the output. With this in mind, we utilized the 2D cylindrically-averaged HI power spectrum (incorporating EoR, noise, and instrumental effects) across the 166--181\,MHz range from the SDC3a test dataset. Subsequently, we generated 50 Gaussian realizations from this spectrum to create a set of HI sky maps. By adding these maps with simulated foreground maps (with instrumental effects), we obtained 50 mock images. For each mock image at a given frequency, we performed reconvolution, PCA subtraction and power spectrum estimation using the same procedure used in processing the SDC3a data, thus getting $\tilde{P} (k_\perp, k_\parallel)$. Then, we estimated the transfer function and its standard deviation from the ratios of the true HI spectrum (provided with the test dataset) and the spectra derived from the mocks. By using those transfer functions, $P$ and its associated statistical errors for SDC3a were finally determined.
%%%%%%%%%%%%%%%%%%%%%%%%%%%%%%%%%%%%%%%%%

\subsection{KORSDC (Kyungjin Ahn, Minji Oh, Jaebeom Kim, David Parkinson, Namuk Lee, Dahee Lee, Minsu Kim, Shinna Kim, Sungwook E. Hong, Se-Heon Oh, Hyeseung Lee, Eunyu Lee, Jae-Young Kim, Junhyun Baek, Jaehong Park, Bong Won Sohn, Hyunwoo Kang)}
\label{KORSDC}
%%{\color{red} What do you mean with "antenna effect"? do you mean the instrumental effects in general, or maybe the Primay beam or the PSF?} 
%%{clarified now}
The team is composed of members of the Korea SKA consortium (Korea SKA), which was established about a decade ago for seeking the governmental participation in the SKA project. We used commonly known schemes on source subtraction, image smoothing and spectral analysis except when they were not sufficient to meet the requirement.

We mainly used the natural-weighted image cube that is believed to preserve the large-scale information better than the uniform-weighted one. The main pipeline is composed of the following sequential steps:
\begin{itemize}
    \item Removal of point sources
    \item Mitigation of the impact of the PSF, the primary beam and the noise
    \item Removal of the foregrounds
    \item Calculation of the power spectrum
\end{itemize}
Throughout our pipeline, we used the central $3 \times 3 \,{\rm deg}^2$ out of the given $9 \times 9\,{\rm deg}^2$ to mitigate the impact of the noise, which is amplified after the primary beam correction as one moves away from the beam center. In the following subsections we describe details of each step. In addition, we utilized the characteristics of the data set that resulted in a fudge factor used to calibrate the amplitude of the power spectrum (section~\ref{KSDC_pk}). Overall, we suffered mostly from the impact by the PSF that seemed to dominate the image even after our point-source and foreground removal.

\subsubsection{Removal of point sources}
\label{KSDC_ps}
We attempted to remove point sources smoothly from the original image data cube. The usual extraction of point sources leave blanks on the image data, but we found that the power spectrum is affected even at the smallest $k$ of the targeted $k$ range of the challenge. We used \texttt{SExtractor}\footnote{https://github.com/astromatic/sextractor} (``Source Extractor'') to detect and remove point sources. In each frequency channel, the \texttt{SExtractor} parameter {\tt BACK SIZE} needs to be defined to determine the background value for each pixel. Following \cite{Bertin96}, we define a new background value for each pixel based on the pixels included in {\tt BACK SIZE}.

\begin{flushleft}
• Non-crowded case = mean \\
• Crowded case = 2.5 $\times$ median - 1.5 $\times$ mean \\
\end{flushleft}

According to \cite{Bertin96}, for crowded regions, the average of the pixel values included in {\tt BACK SIZE} is iterated until it reaches 3\,$\sigma$. If at each iteration the $\sigma$ decreases by 20\% or less, the region is considered non-crowded. Therefore, to detect point sources among irregular and extended signals, it is necessary to consider the overall brightness of extended sources close to point sources. To achieve this, we set the number of pixels included in {\tt BACK SIZE} to 10, and the newly defined background value worked as a photometric zero point for point sources. After separating point sources from irregular and extended sources and conducting visual inspection on residuals, we recombined all frequency channels to recreate the 3D data cube.

\subsubsection{Image into the brightness temperature / primary beam correction}
\label{KSDC_bt}
The provided image cube was in Jy/Beam. Assuming a 2D Gaussian shape with axes of 0.05~deg and 0.04~deg for the synthesized beam, we divided the given flux density by the beam area to obtain the specific intensity $I_{\nu}$ and used the definition of the brightness temperature in the Rayleigh-Jeans regime,
\begin{equation}
T_{b} = \frac{c^2}{2\nu^{2}k_{B}}I_{\nu},
\end{equation}
To implement this, we adopted routines from the {\tt ps\_eor}
\footnote{\url{https://gitlab.com/flomertens/ps_eor/-/blob/master/ps_eor/datacube.py}} package.

We also performed the primary beam correction, which was unavoidable as we used a substantial part of the full FoV. We divided the image data plane by the time-averaged beam provided at each frequency band due to the variation of the beam size over the frequency.

\subsubsection{PSF and thermal noise}
\label{KSDC_psf}
We tried to mitigate the impact of the PSF as follows. Our original strategy was to perform a ``deconvolution'', or to mitigate the PSF effect by taking the division (Fourier-transformed image)/(PSF) in the Fourier ($k$-) plane, following equation (4) of \cite{2023MNRAS.524.3724C}. This is based on the usual convolution theorem. However, we found that this naive approach did not fully remove the PSF but instead left a residual fluctuation which was substantial enough to impact the resulting image and the power spectrum significantly. We also could not fully understand the normalization convention of the PSF. Specifically, summation of the pixel values of image cube for each frequency within the cut area $3 \times 3 \,{\rm deg}^2$ was often negative. This problem was entangled with the thermal noise correction as well. Since the thermal noise could not be removed either in image space or in Fourier space because of unknown random seed, it still remained in the process of PSF correction. When the image with the thermal noise is divided by the PSF, the resulting image (when inverse Fourier-transformed from the $k$-plane) became uncontrollable in its amplitude and stochasticity. We now believe that we should have smoothed the PSF itself to tame the sidelobes before the deconvolution.

Therefore, we opted to simply smooth the image to reduce the impact of the PSF and the thermal noise with a frequency-dependent smoothing kernel. For this, we used the test data set to find an optimal choice of the smoothing kernel. By trial and error, we adopted a 2D Gaussian filter with the smoothing angle $\sigma=48\arcsec$ at the highest frequency band (181\,MHz), with the kernal in the form $\propto \exp[- \theta^2 /2\sigma^2]$. We smoothed the data at each frequency band by scaling $\sigma$ in proportion to the frequency inverse. To implement this, we utilized multidimensional image processing packages in {\sc scipy} ~\citep{2020SciPy-NMeth}.

\subsubsection{Foreground removal}
\label{KSDC_fr}

We applied ICA on each line of sight  (LoS) to the image obtained as described above, which contains diffuse foreground emission, the EoR signal and the mitigated PSF and noise. We did not utilize the intrinsic angular correlation of the diffuse foreground but just treated LoSs independently. We assumed 3 independent foreground components, and 1 extra component to hopefully capture the unwanted spectral features introduced by the PSF and the noise. After obtaining independent component amplitudes for each LoS, we fitted out the 4 components and obtained the residual image cube. We used the FastICA in the {\sc scikit-learn} package for ICA~\citep{2011JMLR...12.2825P}. Unfortunately, many LoSs showed highly irregular frequency spectra even after the data had been processed (Sections~\ref{KSDC_ps}--\ref{KSDC_psf}). We believe that this caused the ICA to struggle in separating out the EoR signal.

\subsubsection{Power spectrum estimation}
\label{KSDC_pk}
For the 2D cylindrical power spectrum estimation, we first divided the image cube into 6 sub-cubes with the frequency interval of 15\,MHz for each as requested by the SDC organisers. Then, we converted angular coordinates (RA, Dec) of each sub-cube into comoving coordinates using the cosmological parameters ($\Omega_m=0.30964$, $\Omega_{\Lambda}=0.69036$ and $H_{0}=100$) specified in the data description. We Fourier-transformed this coarsened data cube, and collected the power for each ($k_{\parallel}$, $k_{\perp}$) bin where $0.05 < k_{\parallel}$, $k_{\perp} < 0.5 \text{Mpc}^{-1}$ with $\Delta k = 0.05 \text{Mpc}^{-1}$.

Note that we applied the PSF normalization fudge factor 1/3 that we empirically found to the resultant power spectrum: we multiplied 1/3 to the power spectrum we obtained above for the final estimation. We obtained this fudge factor from the test data set as follows. To the test data set, which is free from point sources and the diffuse foreground, we applied the primary beam correction, angular smoothing and the ICA with only 1 component (reserved for the PSF effect on a LoS spectrum). When we fitted out the spectra of LOSs and obtained the power spectrum, the power spectrum was found to be about 3 times as high as the true power spectrum in amplitude but otherwise similar in the overall structure. We then assumed that the PSF effect on the actual data cube would be of the similar level even after the inclusion of the point sources and the foreground.

Our estimated power spectrum is found to be about 2 orders of magnitude as large as the true value. We believe that residuals of the point sources and the diffuse foregrounds, combined with the PSF residual, have not been mitigated to the level we intended and dominated the power spectrum.

\subsection{KUSANAGI (S. Yoshiura, T. Ito, T. Minoda, T. T. Takeuchi, T. J. Hayashi, K. Takahashi, H. Shimabukuro, K. Hasegawa, T. Akahori)}

KUSANAGI consisted of members of EoR 
%{\color{red} please expand the acronyms here} 
Science Working Group and Engineering Working Group in the SKA-jp consortium. 
Our approach can be broken down into 4 parts; (1) creating a source list, (2) removing radio sources from the visibilities, (3) statistical foreground removal in image space and (4) power spectrum estimation. We describe each part briefly below. Note that we gathered some publicly available software such as \texttt{CASA}, \texttt{gpr4im} \citep{Soares2022}, \texttt{PyPBSF} \citep{Mohan2015} and {\tt WSClean} \citep{2014MNRAS.444..606O}. For other processes including point source subtraction, statistical foreground removal and power spectrum estimation, we built our software from scratch and using some Python packages (e.g. \texttt{numpy} and \texttt{astropy}).

For creating source lists used for point source removal, we started by converting the visibilities (MS format) to images with \texttt{WSClean} using only 2\,MHz of data centred at each 15\,MHz band. We ignored the effect of the SKA station beam which has significant frequency dependence. To reduce the diffuse emission components, the baselines were limited to the range from 1000\,$\lambda$ to 5000\,$\lambda$. The radio source identification was performed using \texttt{PyPBSF} on the CLEANed images (fits format) and made 6 different source lists at each 15\,MHz interval.

The source lists were used to make modelled visibilities at each channel based on the information of original SKA visibilities (uvfits format) and SKA station beam model. For the modelling we employed a CUDA scheme developed in RTS \citep{2008ISTSP...2..707M}. 
The spectral index of radio sources was set to -0.7. The modelled visibilities is directly subtracted from the SKA visibilities. The residual visibilities data are converted to MS format using \texttt{CASA}. We again make images with the {\tt WSClean} from the residual visibilities at each channel. At this stage, we made the natural weighted dirty map using a limited range of baseline from 30$\lambda$ to 250$\lambda$. The images consists of image cubes of $512 \times 512 \times 150$ grids at each 15\,MHz. Note that we made two sets of images at even and odd time samples. Thus, there were  12 image cubes (6 bands $\times$ 2 time steps). The images were dominated by diffuse emission and some residual bright sources. 

We then applied two different statistical foreground removal methods in the image space. One was the traditional fitting with a 3rd-order polynomial function using the "polyfit" method in \texttt{numpy}. We simply fitted the spectrum at each pixel along the frequency axis for each image cube. After subtracting the fitted values, the residual image cubes were available for power spectrum estimation. Another method was based on Gaussian Process Regression in the image space. There were 3 covariance kernels (RBF, Matern3/2, exponential). The hyper-parameters were optimized by a GPy implemented optimizer. The mean value of the foreground was estimated by using RBF and Merten3/2 kernels at each pixel along the frequency axis and was removed from the image cubes. We therefore have two sets of residual cubes. The results of the polynomial fitting and the GPR were referred to as KUSANAGI-poly and KUSANAGI-GPR,
respectively. 

Finally, we estimated the power spectrum from the residual image cubes. The procedure performed here was motivated by the method described in \cite{mertens20}. The image cubes were converted to gridded visibilities cubes by the FFT. We applied the Blackman‐Harris window function along the frequency axis before the Fourier transform. The cross-power spectrum was evaluated using two different time-step data at each Fourier mode. The median of cylindrical power spectrum value was evaluated at each $k_{\perp}$ and $k_{\parallel}$. To match the SDC3a required format, we evaluated the averaged value in wider $k$-bins (0.05 $\rm{hMpc^{-1}}$). The variance of samples $\sigma^2$ and the number of samples (N) were also calculated in each wider $k$-bin. We submitted $\sigma/\sqrt{\rm N}$ as the error on each bin of the power spectrum. Even after the foreground removal, the residual had significant power, especially at the foreground wedge \citep{Morales12}. Following the foreground avoidance approach, we removed the k-bin in the foreground wedge assuming the maximum contamination from 10 degrees from the field centre. 

We thank the JP-SRC funded by SKAJ and cooperating by Kumamoto University and Nagoya University for providing computational resources used in our analysis.

%%%%%%%%%%%%%%%%%%%%%%
\subsection{Nottingham-Imperial (Emma Chapman, Luke Conaboy, Jennifer Feron, Carina Norregaard, Jonathan Pritchard)}

We started from the visibility files and used the {\tt OSKAR} visibilities simulation package\footnote{\url{https://github.com/OxfordSKA/OSKAR}} to remove the bright point sources from the data set. This was done by aligning the brightest point sources with the GLEAM catalogue and using this bright point source map to feed into {\tt OSKAR} to remove them.

Using the RA, Dec, and minimum flux of the point sources (100~mJy at 151\,MHz) located within the data cubes, we queried the GLEAM catalogue \citep{2017MNRAS.464.1146H} in VizieR\footnote{\url{https://vizier.cds.unistra.fr/viz-bin/VizieR-3?-source=VIII/100/gleamegc}} to extract an approximate catalogue of corresponding point sources, which we then combined into an {\tt OSKAR} Sky Model. We used {\tt OSKAR} to simulate visibilities over the entire frequency range, in steps of ${\rm d}\nu = 0.1~{\rm MHz}$. Each simulated observation was centred on ${\rm RA}=0^\circ$ and ${\rm Dec}=-30^\circ$, had a total length of 4~hours, a 10~second integration time and used the provided telescope model. We directly subtracted these simulated visibilities from the data challenge visibilities to produce a set of point-source-subtracted visibilities, which formed the basis for the rest of our analysis. 

We next applied the FastICA blind source separation (BSS) technique to remove the remaining diffuse foregrounds \citep{Chapman12}. This technique requires careful choice of the number of components and the frequency bins over which to apply FastICA, in order to model diffuse foregrounds while avoiding over fitting. After application of FastICA to obtain diffuse foreground subtracted residuals, Pearson correlation coefficients were used to compare the similarity of the residual data cube to the foreground model and the bright source removed data cube. By first varying the size of the frequency cube and then the number of components we could find the minimum correlation. We found a binning of 7.5\,MHz allowed the best foreground removal and easily allowed the power spectrum to be made in 15\,MHz bins, as was the required output. Different component numbers, ranging from 4-8 components, were obtained for each frequency bin in this way. 

For the submitted analysis, the bright source subtracted data cube was first cut to be 1024 $\times$ 1024 instead of the full 2048 $\times$ 2048 so only the region within the primary beam was present. This was intended to negate some of the side lobe effects and improve FastICA performance. The data was then split into 7.5\,MHz sections and each section was put through FastICA using the optimum component number. We also explored imposing a fixed resolution at all frequencies, by convolving the frequency sections with a beam of fixed resolution, before applying FastICA. This has been suggested as a way of allowing FastICA to better identify foreground structure. This did not appear to give significant improvement, so ultimately we chose not to sacrifice resolution in this way.

After converting the residual data cubes from Jy/beam to Kelvin, using standard relations, the {\tt tools21cm}\footnote{\url{https://tools21cm.readthedocs.io/}} python package was used to calculate the 2D power spectrum for each 7.5\,MHz section of data (12 total with corresponding error 2D power spectrum). The $k_{\parallel}$ and $k_{\perp}$ bins given by the SKA were fed directly into the calculation for the power spectrum. Uncertainties were calculated based on a thermal noise model converted to a 2D power spectrum added to an estimate of sample variance.

Our work was carried out on the University of Nottingham HPC Augusta. {\tt OSKAR} simulation of visibilities made use of GPUs for acceleration, while FastICA was run on CPUs. 

We plan to further develop this method by optimising the selection of the number of FastICA components in an automated way for blind data where the truth is not known. Much of the challenge for our analysis lay in accurate removal of point sources.
%%%%%%%%%%%%%%%%%%

\subsection{Pisano Galaxy Moppers (A. Nasirudin, S. Murray, A. Mesinger)}
Our team utilizes the foreground avoidance method in a 21\,cm inference setting to ``recover’’ the underlying cosmic signal. We use \texttt{21cmmc} \citep{park2019inferring} to generate 21\,cm lightcones and perform the Bayesian inference, while the data reduction and likelihood computation is done on-the-fly using \texttt{py21cmmc-fg}, the plug-in code to \texttt{21cmmc}. To maximize the time spent on sampling the parameter space, we have made several assumptions and simplifications:
\begin{itemize}
\item In both data and model, we completely forgo gridding of the visibilities, and instead, calculate the delay spectrum. Once we have the final mean and uncertainty for the frequency bins, we convert to the power spectrum.
\item We split the visibility data into odd/even timesteps, calculate the cross-(delay) power between them, and average all power in each frequency bin.
\item We only model the sky over the central 4x4 degree FoV. To generate our sky model, we undertake similar steps used to generate the SDC3a data described in Section \ref{sec:skymodel} except we do not use T-RECS for the faint extragalactic source population. Instead, we use a power-law relation with parameters from \citep{2011A&A...535A..38I}.
\end{itemize}

We gratefully acknowledge computational resources of the Center for High Performance Computing (CHPC) at Scuola Normale Superiore (SNS) and the Pleiadi infrastructure by INAF.

\subsection{REACTOR (Shulei Ni, Huaxi Chen, Hao Chen, Xuejian Jiang)}
We have developed an innovative algorithm for deconvolution of astronomical images, namely {\tt{PI-AstroDeconv}}, integrating physical information with semi-supervised learning techniques~\citep{ni2024pi}. Its design philosophy has been further developed based on our previous research achievements~\citep{Ni:2022kxn}.

During the development of the algorithm, we first performed beam removal operations on the astronomical images to eliminate the impact of the beam on the images. Subsequently, we applied PCA to the images after beam removal for foreground subtraction, aiming to enhance the image quality. Next, we utilized the {\tt{astropy.FlatLambdaCDM}} and {\tt{scipy.states}} libraries to calculate the dimensions (dims) of the 3D cube and the average value within each two-dimensional region after binning, respectively. 

\subsubsection{Beam Effects on Foreground Removal}
In our previous research, we noted that intricate beam effects had a considerable impact on the outcomes of PCA~\citep{Ni:2022kxn}. To emulate full-sky neutral hydrogen observational data, we utilized the CRIME simulator, superimposing multiple components before convolving them with two distinct beam types: an ideal Gaussian beam and the cosine beam employed by the MeerKAT telescope \citep{Matshawule:2020fjz}. The results demonstrated that PCA could successfully remove foreground components affected by the Gaussian beam, yet it struggled to eradicate those influenced by the cosine beam. Nevertheless, the integration of the UNet neural network significantly enhanced performance.

Based on the aforementioned conclusions, we conducted further validation. We first apply the UNet model to the observational data for de-convolution to counteract the beam smearing effect, with the input being the sum of the convolved foreground and signal, and the label being the sum of the original, unconvolved foreground and signal. Following this, we use PCA to process the data and eliminate the foreground components. The results indicated that this method remains effective. It can be represented by the following formula:
\begin{equation}
    \begin{split}
    &\text{UNet}\{\text{PCA}_{\rm res}[\text{Beam}_{\rm sm}(\rm{FG}+\rm{HI})]\} \approx  \rm{HI} \Leftrightarrow \\
    &\text{PCA}_{\rm res}\{\text{UNet}[\text{Beam}_{\rm sm}(\rm{FG}+\rm{HI})]\} \approx  \rm{HI},
    \end{split}
\end{equation}
where $\text{PCA}_{\rm res}$ denotes the residual values after $\text{PCA}$ processing, UNet refers to the results processed through the $\text{UNet}$ network, $\text{Beam}_{\rm sm}$ indicates the convolutional beam operation applied to the signal, $FG$ represents the cumulative sum of the individual component foregrounds, and $HI$ stands for the neutral hydrogen signal. 

It is evident from the aforementioned formula that the order of applying PCA followed by the UNet network or vice versa does not affect the final analytical outcome. This demonstrates that both sequences of processing are capable of effectively removing the foreground signals from the data, thus achieving the desired analytical objectives. Based on this, we propose a semi-supervised learning algorithm that integrates physical information, aimed at mitigating the effects of beam patterns in observational data.

\subsubsection{PI-AstroDeconv Model}
The pivotal innovation of the {\tt PI-AstroDeconv}~\citep{ni2024pi} netowrk lies in the integration of physical information from astronomical observations—specifically, the telescope PSF—into the model training process, as depicted in Figure 1 of \citet{ni2024pi}. Within the network, the intermediate prediction layer acts as the output layer, corresponding to the ``Prediction'' section shown in Figure 1 of \citet{ni2024pi}. The training objective of this network is to achieve complete identity between the input and output. In this context, the convolution of the ``Prediction'' layer with the PSF represents the observed data, while the UNet is responsible for pixel-level restoration. Thus, we can deduce that the ``Prediction'' output from the intermediate layer equates to the observation image after the removal of convolutional effects. Ultimately, we trained each frequency band separately, resulting in a 3D data cube.

It is worth noting that in our research, we chose to perform deconvolution on the original images, meaning that both the images and the PSF in the training dataset have dimensions of $2048 \times 2048$ pixels. This choice leads to a significant slowdown in the training speed due to the mathematical convolution operation added in the last layer of the network. To effectively address this issue, we innovatively employed FFT technology to redesign the mathematical convolution process, thereby significantly improving the training efficiency.

Building on this, we further mitigated the impact of the beam effect and employed the PCA algorithm to perform foreground subtraction on the network. Through this process, we removed six major components, obtaining the residual image after foreground removal, which was then used to calculate the cylindrical (2D) power spectrum. 

\subsubsection{Power Spectrum Estimation}
Following the standards set by SDC3a for simulated data description and parameter configuration, we selected the central $512 \times 512$ region of data for each frequency band for analysis. To estimate the cylindrical power spectrum, we first subdivided the image cube into six sub-cubes with a frequency bandwidth separation of 15\,MHz. We then compute the {\tt{box\_dims}} size using the {\tt{FlatLambdaCDM}} model from the {\tt{Astropy}} library, followed by applying the {\tt{binned\_statistic\_2d}} function from the {\tt{Scipy}} library to perform binned statistical analysis on the two-dimensional data. Subsequently, we subdivide the data according to the bins recommended by SDC3a and perform specific statistical operations within each partition to generate the corresponding cylindrical power spectra. As for the error calculation step, we follow the same process but substitute the standard deviation for the mean to obtain an error estimate.

%%%%%%%%%%%%%%%%%%%%%%%%
\subsection{Shuimu-Tianlai (Shifan Zuo, Kangning Diao, Richard Grumitt, Yi Mao, Xuelei Chen, Furen Deng, Yan Gong, Yuer Jiang, Yichao Li, Yingfeng Liu, Qingbo Ma, Hayato Shimabukuro, Tian-Yang Sun, Qiao Wang, Xin Wang, Yu-Xin Wang, Jun-Qing Xia, Yidong Xu, Ruiqing Yan, Ye-Peng Yan, Zongyao Yin, Kaifeng Yu, Xianchuan Yu, Bin Yue, Li Zhang, Xin Zhang, Xingchen Zhou)}
The team used a new method called oriented singular value decomposition (O-SVD) \citep{Zeng2020} for the foreground subtraction task.

The O-SVD of a third-order tensor $\mathcal{A} \in \mathbb{R}^{I_{1} \times I_{2} \times I_{3}}$ with rank $R_{3} = \text{rank}_{3}(\mathcal{A})$ is
\begin{equation*}
  \mathcal{A} = (\mathcal{U} *_{3} \mathcal{S} *_{3} \mathcal{V}) \times_{3} U^{(3)},
\end{equation*}
where the orthogonal matrix $U^{(3)} \in \mathbb{R}^{I_{3} \times I_{3}}$, three tensors $\mathcal{U} \in \mathbb{R}^{I_{1} \times I_{1} \times I_{3}}$, $\mathcal{S} \in \mathbb{R}^{I_{1} \times I_{2} \times I_{3}}$, $\mathcal{V} \in \mathbb{R}^{I_{2} \times I_{2} \times I_{3}}$ such that\\
(1) $\mathcal{U}(:, :, i)$, $\mathcal{V}(:, :, i)$ are orthogonal and $\mathcal{S}(:, :, i)$ is a non-negative diagonal matrix for $i = 1, \cdots, R_{3}$;\\
(2) $\mathcal{U}(:, :, i)$, $\mathcal{V}(:, :, i)$ and $\mathcal{S}(:, :, i)$ are all zero matrices for $i = R_{3} + 1, \cdots, I_{3}$.\\
The diagonal elements $s_{jji}$ of each frontal slice of $\mathcal{S}$ are called the singular values of the pair $(\mathcal{A}, \mathcal{S})$.

The given image cube can be taken as a third-order tensor, therefore the O-SVD method can be applied.
Similar to the matrix SVD method, we can order all the singular values in descending order and truncate the largest SVD modes as foregrounds to be subtracted from the data. 

The data we have used is the uniform weighted image cube data, which has pixel size of $16 \times 16$ arcsec and number of pixels in RA/Dec of $2048 \times 2048$, and is produced by using the {\tt WSClean} task followed by Gaussian tapering and a multi-scale deconvolution. The frequency coverage of the dataset is 106 -- 196\,MHz. According to the submission requirement, the full frequency range of 90\,MHz is divided into 15\,MHz intervals, and for each sub-band a power spectrum is computed. To limit noise, only the central $4 \times 4$ degrees are used. For such an image cube, we first use the \texttt{radio\_beam}\footnote{\url{https://pypi.org/project/radio-beam/}} package to convert the unit of the data from Jy/beam to K with the BMAJ and BMIN in the FITS file header, and then use the FlatLambdaCDM model in \texttt{astropy.cosmology}\footnote{\url{https://docs.astropy.org/en/stable/cosmology/index.html}} to convert the angle and frequency scale to physical size in unit of Mpc. Such an image cube can then be foreground subtracted with the O-SVD method and used for cylindrical 2D power spectrum estimation. We have used the code in \texttt{tools21cm}\footnote{\url{https://github.com/sambit-giri/tools21cm/blob/master/src/tools21cm/power_spectrum.py}} to compute the cylindrical power spectrum, and the $1 \sigma$ error of the power spectrum was given by the standard deviation of the values that fall into a specific $(k_\perp, k_\parallel)$ bin. We did not use the technique of transfer function to compensate for signal loss by the foreground subtraction method, so we find that the power in small $k$-modes of our submitted results is lower than the true power, indicating there is indeed some signal loss at large scales. The complete code used for this analysis was open-sourced in zenodo\footnote{\url{https://zenodo.org/records/10124117}} and Github\footnote{\url{https://github.com/zuoshifan/ska_sdc3a_pipeline}}.

One of the primary challenges in implementing the method lies in determining the optimal number of foreground modes to be truncated, as the singular values plot does not exhibit an obvious gap for reference. Currently, this number is determined through visual comparison of the residual images with test images that lack foreground. However, this approach is suboptimal. To address this limitation, we aim to establish more effective methods for determining the number of foreground modes. We plan to explore model selection techniques such as the Akaike Information Criterion (AIC), the Bayes Information Criterion (BIC), Generalized Cross-Validation (GCV), among others, to derive a more objective and reliable criterion.

\subsection{SKACH (M. Bianco, S. K. Giri, R. Sharma, S. Krishna, T. Chen, C. Finlay, V. Nistane, P. Denzel, M. De Santis, H. Ghorbel)}\label{sec:team_skach}

The team comprised cosmologists, radio interferometry experts and data scientists members of the SKA Switzerland consortium (SKACH\footnote{\url{https://skach.org}}). Our approach consisted of three preprocessing steps followed by a deep-learning framework that recovers the 21\,cm signal from tomographic data with residual foreground contamination and noise.

For the preprocessing, we first employed the composite GLEAM and LoBES catalogue to remove the brightest extragalactic compact sources. We used the available catalogue and computed the corresponding raw visibilities, $V_\mathrm{exgf}$, with {\tt OSKAR}, then subtracted the visibilities from the data challenge visibilities, $V_\mathrm{SDC3a}$, to obtain, $V_\mathrm{res1} = V_\mathrm{SDC3a} - V_\mathrm{exgf}$. Once they were removed, we created dirty images at each observed frequency with {\tt WSClean}\footnote{We used the same parameters as described in the SDC3a data product descriptions}. We employed \texttt{PyPBSF} on each image to position faint sources in the sky and model their flux distribution with an ellipsoid. The resulting catalogue is used to compute the raw visibilities from which we subtract to the previous step $V_\mathrm{res2} = V_\mathrm{res1} - V_\mathrm{PyPBSF}$. We repeat this process until we reach the flux limit for unresolved sources of $100\, \rm mJy$. To finalize our first pre-processing step, from the residual visibilities, $V_\mathrm{res2}$, we apply uv-sampling filtering for points with $U>200\,\lambda_{\rm obs}$, where $\lambda_\mathrm{obs}$ correspond to the observed wavelength and employ the resulting visibilities to create a tomographic dirty image. The second step consists of a spatial-domain decomposition of the previously mentioned tomographic data. For each pixel in the image, we fit a polynomial function along the line of sight to remove the large contribution of the residual foreground emission, thus producing a dirty image, $I_\mathrm{res}$, with shape $1024^2$ and angular resolution of $14''$ in which ideally most of the foreground contamination is removed.

We employed a U-shaped convolutional neural network based on a modified version of the \texttt{SegU-Net} framework presented by \cite{Bianco2021, Bianco2023}. 
This network was meant for binary segmentation of neutral and ionized regions from 2D tomographic images of the 21\,cm signal during the late stage of reionization, $z < 11$ ($\nu_\mathrm{obs} > 118 \, \rm MHz$). We modified the original architecture to process the tomographic data, and it is similar, but for 3D inputs with shape $\left(128, 128, 16\right)$, to the \texttt{RecU-Net} architecture presented in \cite{Bianco2024}. 

We created a dataset of approximately 20 simulations with different astrophysical parameters and initial conditions to train and test the network. We employed the {\tt 21cmFAST} semi-numerical code \citep{2011MNRAS.411..955M} to simulate the 21\,cm signal and followed the description of the challenge to create a mock observation. We simulated the Galactic synchrotron foreground signal following the method in \cite{Choudhuri2014foreground} and employed the {\tt OSKAR} code \citep{oskar} to simulate the raw visibilities for a 4-hour observation, including the ionospheric effect with the \texttt{ARatmospy} \citep{2015OExpr..2333335S} code. We applied the same time-evolving phase screen for all the data. Additionally, we included systematic noise for the same observational length and decreased the noise level by a factor of 250. With the same approach, we applied direction independent gain error following the model by \cite{Wang2024fr}. In this training dataset, we do not include any foreground contamination as we assume that the preprocessing steps will be able to almost completely remove the point sources and synchrotron galactic emission.

The network was trained for approximately 250 epochs and only for simulated data within $166$ and 181\,MHz. In our previous studies, we demonstrated that the neural network shows an improved performance at frequencies where the 21\,cm standard deviation is maximized. For this reason, we decided to focus our effort on this frequency range. Moreover, the limited size of the dataset and the frequency coverage were limited by computational resources. Because of the limited GPU memory, we could not train a network that manages an input with a mesh size of $1024$. Instead, the neural network trains and predicts only a small portion of the entire image, with a size of $128$, so that we can later patch them together. We apply the trained network on the residual dirty image from our preprocessing steps, $I_\mathrm{res}$. We select these $128^2$ regions to overlap with each other, such that we can calculate the average and standard deviation of the predicted 21\,cm signal based on the number of overlaps in each pixel. We then calculated the 2D power spectra for the corresponding frequency range with the \texttt{tools21cm} software \citep[][]{giri2020tools21cm}. The resulting average and standard deviation were then submitted as the final result and error, respectively.

\subsection{SROT (Akash Kulkarni, Nirmala S R, Basawaraj)}
SROT (Space  Radio  Observation 
 and  Testing) studies the signals from the cosmos through its in-house design of antenna systems and code pipeline design. We used GPR to solve the proposed challenge. 
As presented in \cite{Soares_2021} and 
\cite{2021MNRAS.500.2264H} it is evident that GPR is a better-performing solution at lower frequencies.  \cite{Soares_2021} express that it outperforms the PCA method. Hence, we decided to use the \texttt{gpr4im} python package \footnote{\url{https://github.com/paulassoares/gpr4im}}. The code of \texttt{gpr4im} was written for MeerKAT's HI observation data.

We utilized the \texttt{gpr4im} package and made needful changes to achieve the results.  We used image cube data in both \emph{natural} weighted and  \emph{uniform} weighted but presented the results obtained by \emph{natural} weighted data. The \texttt{gpr4im} package supports fetching the cylindrical power spectrum, as requested by the challenge. The result of these changes and the adjustment to run the whole code within the resources available, i.e., 32 cores and 125\,GB of RAM, is presented in the code\footnote{\url{https://github.com/AkashRadio/SKA_SDC3}}.

The SROT team acknowledges the Spanish Prototype of an SRC (SPSRC) service and support funded by the Spanish Ministry of Science, Innovation and Universities, by the Regional Government of Andalusia, by the European Regional Development Funds and by the European Union NextGenerationEU/PRTR. %The SPSRC acknowledges financial support from the State Agency for Research of the Spanish MCIU through the "Center of Excellence Severo Ochoa" award to the Instituto de Astrofísica de Andalucía (SEV-2017-0709) and from the grant CEX2021-001131-S funded by MCIN/AEI/ 10.13039/501100011033 \cite{SPSRC_cite}

\subsection{Wizards of Oz 3D (CH Jordan, JH Cook, JB Line, D Null, C Nunhokee, B Pindor, A Selvaraj, CM Trott)}\label{sec:team_template1}
The team was composed of Australia-based members of the MWA EoR project. We chose to approach the SDC3a challenge with the same methodology as used by the team for the MWA, in an attempt to explore how well our methods were applicable to these data. As such, we took the same calibration, and compact source subtraction approach, and did not attempt to undertake any sophisticated foreground fitting. %It is worth noting that we used the SDC3a test dataset as a mean to calibrate our results, and therefore our normalization was affected by the systematic effect described in Sec.\ref{sec:systematics}, we were unable to re-submit our results after the error was found in the test dataset, due to loss of team members. 
The test dataset was used to define the normalisation of our results, but the team was unable to re-submit our results after the issue described in Sec. \ref{sec:systematics} was discovered, due to loss of team members. As such, our results retain the bias inherent in the test dataset.

The team used a custom version of \texttt{mwa\_hyperdrive}, which is the primary calibration software for MWA \citep{tingay_2013,wayth_2018} EoR data in Australia \citep{hyperdrive_2024}. \texttt{Hyperdrive} is open source and licensed under the Mozilla Public License version 2.0. We made a number of improvements and accommodations to \texttt{hyperdrive} for SDC3a:
\begin{itemize}
\item The ability to read, calibrate and write single-polarisation data (typically deal with full-polarisation data, so Jones matrices are assumed as the “unit” of a visibility throughout the code.  Using only a single polarisation meant that calibration would always fail, because all our Jones matrices were singular.)
\item The ability to remove precessed UVWs uvfits; measurement set formats expect their UVWs to be in the J2000 frame, but {\tt OSKAR} appears to only write UVWs in the observation frame.
\item Inclusion of an “Airy disk” beam model, which was used for the SDC3a data
\item A small offset between the supplied UVWs and our generated UVWs was noticed.  We are not exactly sure how this manifested, but we were able to robustly match the supplied UVWs when using an additional time offset.
\end{itemize}

After creating our 12 timesteps, full bandwidth files and finding the time offset according to the UVWs, we investigated calibration.  We found no gain errors, so we abandoned calibration.  We also found that any ionospheric offsets that were present were too small to be noticeable.  Instead we focused most of our time on improving our sky model to improve the quality of the sky-model-subtracted visibilities. We found that the SDC3a-supplied LoBES sky model did not work as well as another version of LoBES we managed to locate.  After stripping LoBES sources that did not appear to be in the SDC3a data, we ran the aegean source finder \citep{aegean} on a cleaned image to obtain a sky model for the T-RECS sources.  We then iterated on improving the sky model until our subtracted visibilities looked good in images as well as power spectra.

The power spectrum estimation methodology followed that used by the CHIPS software for MWA data \citep{trott_2016}, with updated parameters relevant for SKA. The data were split into even/odd timesteps for power spectrum estimation, and to remove noise power bias. The calibrated and subtracted visibilities were gridded onto the uv-plane using a Blackman-Harris gridding kernel matched to the instrument FoV, along with a separate weights grid. For each uv-cell, a 4th-order polynomial was fitted to the real and imaginary components of the spectrum and subtracted, in order to reduce residual diffuse emission. A 4th-order polynomial has sufficiently long coherence lengths to not affect the cosmological signal. The cubes were then split into an individual cube for each redshift range. The data were Fourier Transformed along the spectral direction, after weighting with a Blackman-Harris window.  Visibility cubes were normalised by their weights, and then cylindrically-averaged and squared to produce the final power spectra. Noise was calculated using the difference set of visibilities. Data analysis was performed using resources of the Pawsey Supercomputing Research Centre.

\section{Assessment of results} \label{sec:analysis}
\begin{table}
\caption{List of teams that employed foreground avoidance as part of their strategy, and percentage of data submitted per frequency interval (lowest frequency from left to right) \label{tab:avoidance}}
\begin{center}
\begin{tabular}{l l l l l l l}
%\hline
\hline
$\nu_{\rm min}$[MHz]&106&121&136&151&166&181\\
$\nu_{\rm max}$[MHz]&121&136&151&166&181&196\\
\hline
%Team&&&&&&\\
%\hline
Cantabrigians&64&60&62&64&65&70\\
DOTSS-21cm\_Avoidance&66&67&68&70&70&71\\
HAMSTER&56&58&65&60&56&55\\
KUSANAGI-poly&19&20&25&25&27&30\\
KUSANAGI-GPR&19&20&25&25&27&30\\          
\hline
\end{tabular}
\end{center}
\end{table}

\begin{table}
\caption{SDC3a Leaderboard, based on the score of eq. \ref{score}. \label{tab:leaderboad}}
\begin{center}
\begin{tabular}{l l l}
\hline

Rank&Team&Score\\
\hline

1&DOTSS-21cm Advanced ML-GPR&240226\\
2&DOTSS-21cm ML-GPR&228445\\
3&HIMALAYA&134752\\
4&DOTSS-21cm Avoidance&109567\\
5&Shuimu-Tianlai&98128\\
6&ERWA&71885\\
7&Wizards of Oz 3D&59513\\
8&Akashanga&40224\\
9&SKACH&37202\\
10&Hausos&26315\\
11&REACTOR&\\
12&KUSANAGI-poly&\\
13&Cantabrigians&\\
14&Nottingham-Imperial&\\
15&KUSANAGI-GPR&\\
16&Pisano Galaxy Moppers&\\
17&HAMSTER&\\
18&Foregrounds-FRIENDS&\\
19&KORSDC&\\
20&SROT&\\
\hline
\end{tabular}
\end{center}
\end{table}

\begin{table}
\caption{Percentage of the SDC3a score achieved within each of the six frequency bins} 
\label{tab:score_freq}
\begin{center}
\begin{tabular}{l l l l l l l}
\hline
$\nu_{\rm min}$[MHz]&106&121&136&151&166&181\\
$\nu_{\rm max}$[MHz]&121&136&151&166&181&196\\
\hline
Akashganga&1&6&39&15&17&19\\
Cantabrigians&3&6&0&1&7&80\\
DOTSS-21cm&1&5&7&19&24&42\\
ERWA&0&24&24&27&15&6\\
Foregrounds-FRIENDS&0&0&0&0&0&99\\
HAMSTER&2&6&15&18&22&33\\
Hausos&0&3&79&12&4&0\\
HIMALAYA&0&30&8&15&16&27\\
KORSDC&0&0&9&16&38&35\\
KUSANAGI&5&0&0&0&44&50\\
Nottingham-Imperial&0&0&0&4&95&0\\
Pisano Galaxy Moppers&1&4&10&19&28&35\\
REACTOR&7&18&18&18&18&18\\
SKACH&0&0&56&17&7&18\\
SROT&16&16&16&16&16&16\\
Shuimu-Tianlai&0&2&9&12&28&45\\
Wizards of Oz 3D&1&0&7&26&29&35\\
\hline
\end{tabular}
\end{center}
\end{table}

\begin{figure*}
\includegraphics[width=17cm]{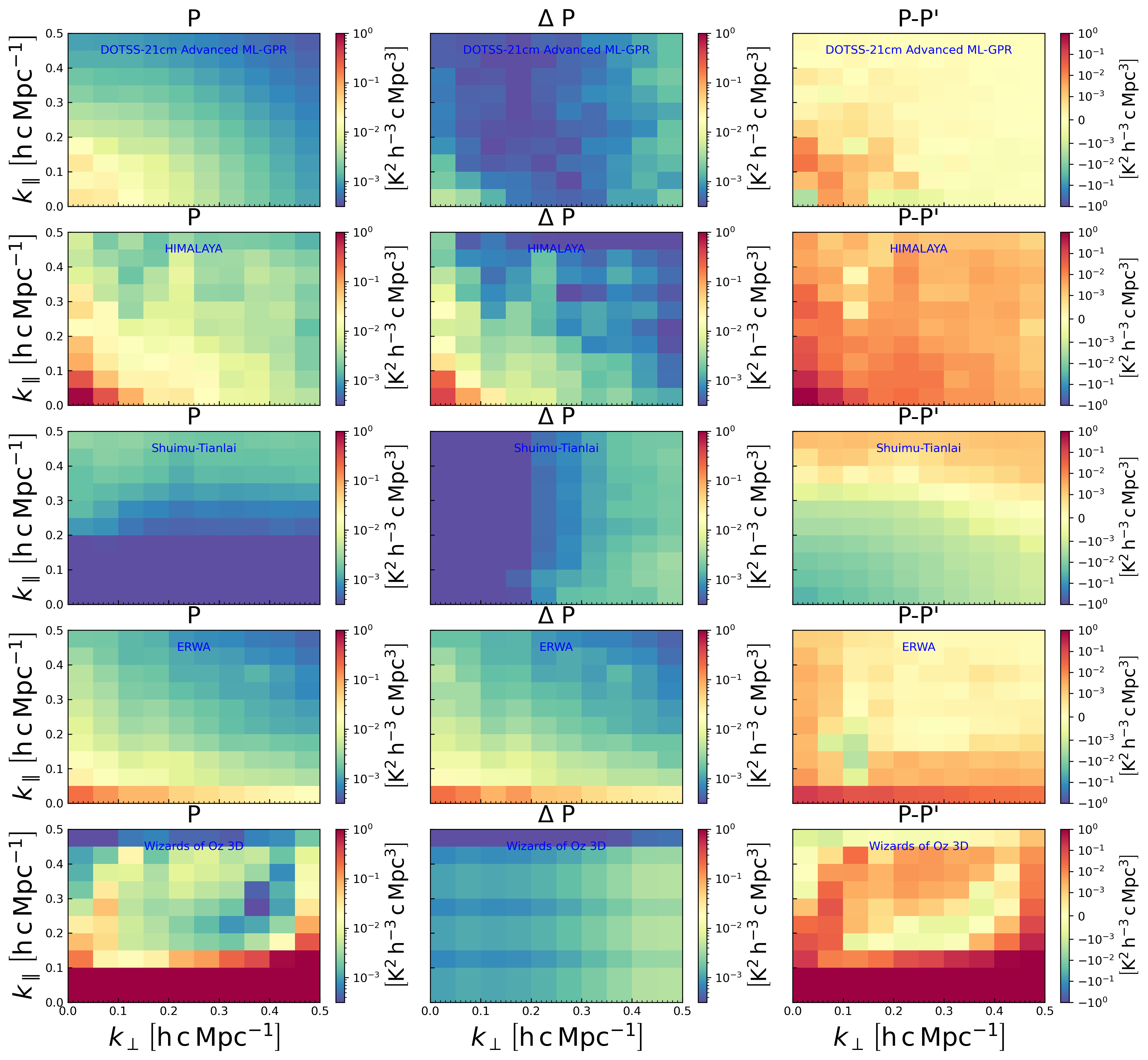} 
\caption{SDC3a submissions for the 151--166\,MHz frequency and assessment for the top 5 teams (with only one result per team) in the SDC3a leaderboard. Left: power spectrum submission $P$, to be compared with the corresponding true power spectrum $P'$ in Fig. \ref{fig:truth}. Centre: submitted 1\,$\sigma$ error bar $\Delta P$. Right: True error committed, $P-P'$. \label{fig:fullPS}}
\end{figure*}

\begin{figure*}
\includegraphics[width=8.cm]{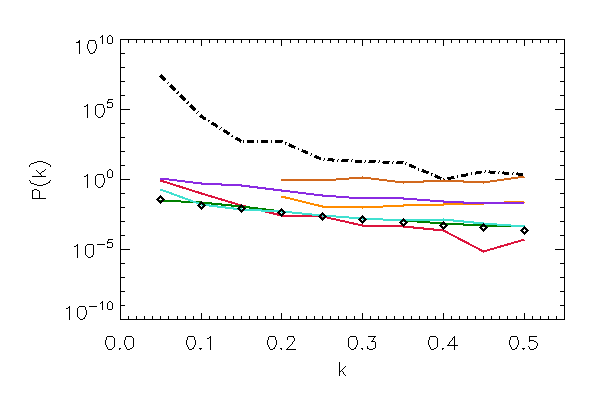}             
\includegraphics[width=8.cm]{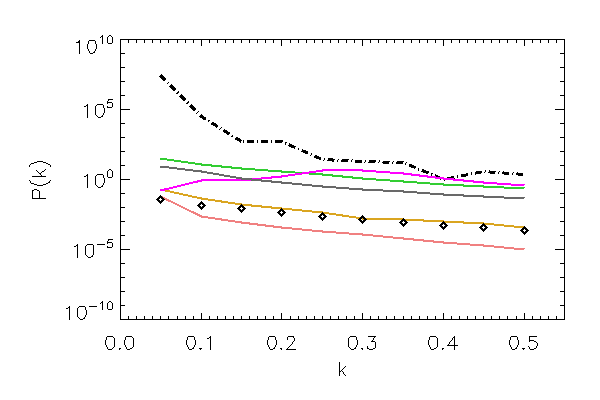}             \includegraphics[width=8.cm]{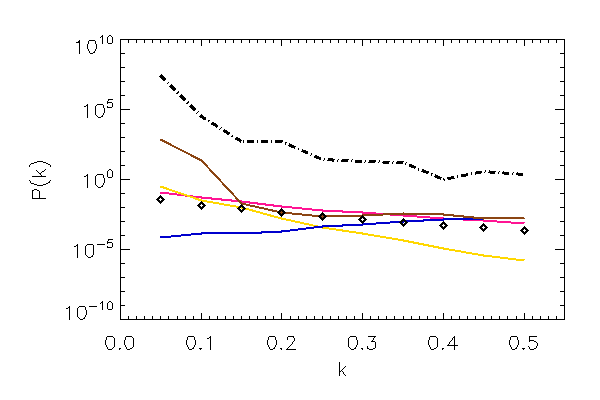}     
\includegraphics[width=8.cm]{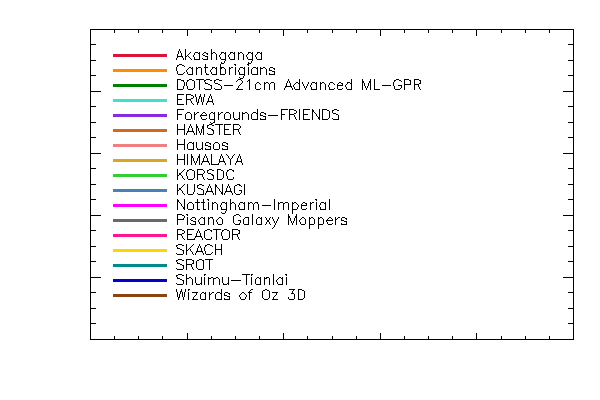}   
\caption{Diagonal terms ($k_\parallel=k_\bot$) of $P(k_\parallel,k_\bot)$ for the 151--166\,MHz frequency range. Black diamonds: true EoR power spectrum $P'$; black dot-dashed lines: true total power spectrum, including foregrounds. Coloured lines: EoR power spectra $P$ recovered by teams, spread into three panels for clarity. \label{fig:allsub}}
\end{figure*}

\begin{figure}
\includegraphics[width=8.cm]{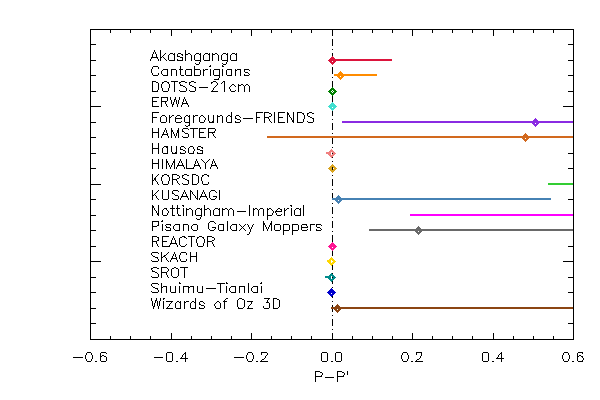}   
\includegraphics[width=8.cm]{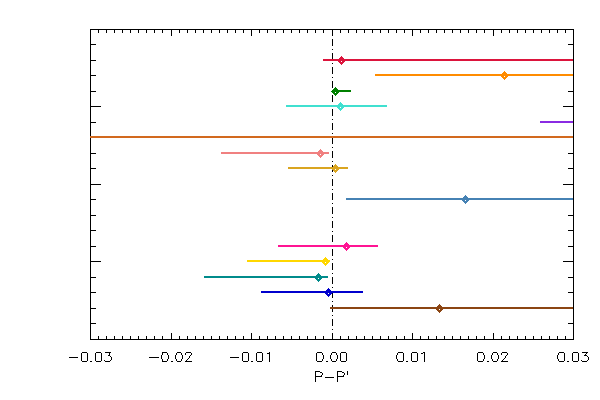}   
\caption{Median (diamonds) and 16\%--84\% percentiles (bar) of the error distribution $P-P'$ over all frequencies and scales for all the teams, each represented by a different colour. The value on the y axis has no meaning and it is used to displays the teams in alphabetical order. \emph{Top:} expanded error range to include all teams; \emph{bottom:} zoom-in around $P-P'=0$.\label{fig:biases}}
\end{figure}

\begin{figure*}
\includegraphics[width=8.cm]{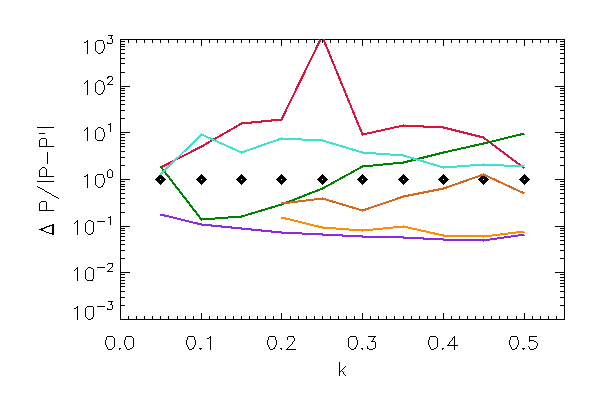}             \includegraphics[width=8.cm]{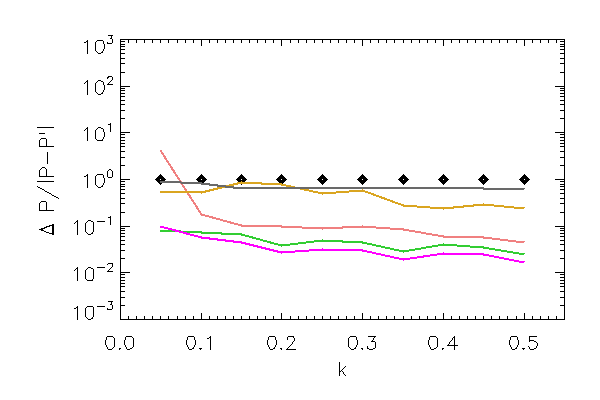}             \includegraphics[width=8.cm]{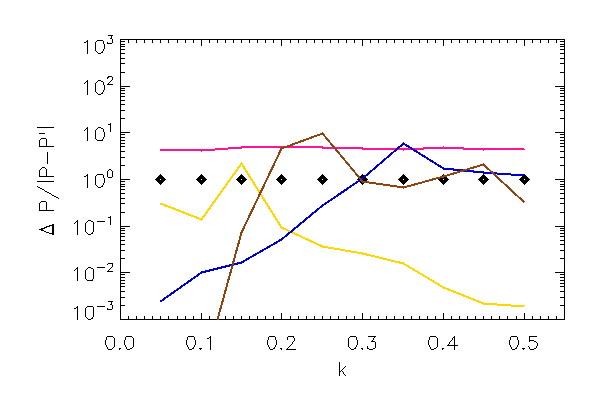}     
\includegraphics[width=8.cm]{figures/Legend.png}   

\caption{Diagonal terms ($k_\parallel=k_\bot$) of $\Delta P(k_\parallel,k_\bot)/|P(k_\parallel,k_\bot)-P'(k_\parallel,k_\bot)|$ for the 151--166\,MHz frequency range.  Coloured lines: EoR power spectra recovered by teams, spread into three panels for clarity. Black diamonds: fiducial value corresponding to perfectly accurate error bars; value below and above the diamonds indicate that the error bars have been underestimated and overestimated, respectively. \label{fig:allerr}}
\end{figure*}

\begin{figure}
\includegraphics[width=8.cm]{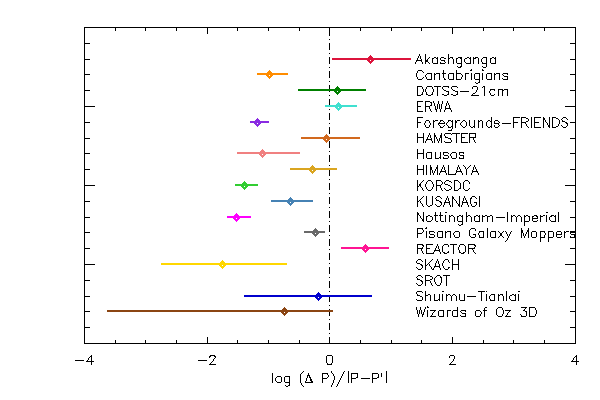}   
\caption{Peak (diamonds) and width at half maximum (bar) of the distribution of $\log(\Delta P/|P-P'|)$ over all frequencies and scales for all the teams. The value on the y axis has no meaning and it is used to displays the teams in alphabetical order, with the colours as in the legend of Figure \ref{fig:allsub}. Points to the left (right) of the vertical line corresponds to underestimated (overestimated) error bars. \label{fig:errorbars}}
\end{figure}

In this section we assess the performance of the participating teams by means of different metrics, starting from the SDC3a score (\ref{sec:leaderboard}) and then moving to power spectrum residuals (\ref{sec:residuals}) and error bar accuracy (\ref{sec:errorbar}). 

\subsection{SDC3a score and leaderboard} \label{sec:leaderboard}
The SDC3a results submission consists of a list of power spectrum values with error bars $P_i(k_\parallel,k_\bot)\pm \Delta P_i(k_\parallel,k_\bot)$ where $i=1,6$ represents the power spectra covering the 90\,MHz frequency interval in 15\,MHz slices, and $(k_\parallel,k_\bot)$ are the bins in the $k$-modes parallel and perpendicular to the line of sight. 

The SDC3a score has been designed to take into account the accuracy of both the EoR power spectrum $P$ and of the error bars $\Delta P$, over the full range of scales $k =0.05$--0.5 and frequencies $\nu=106-196$\,MHz. Since we are neglecting correlation between bins in the error characterization, all power spectrum values are treated as independent. To simplify the notation, we can introduce a new index $j$ that runs through all power spectra and $k$ bins. In this notation, $P_j\pm \Delta P_j$ is the recovered result and $P'_j$ is the true input (shown in  Fig. \ref{fig:truth}). 

In the Gaussian approximation, for each $j$ we can compute the probability of the true value $P'_j$ given the measurement $P_j\pm \Delta P_j$ as
\begin{equation}
\Pr(P'_j)=\frac{1}{\sqrt{2\pi}\Delta P_j}\exp[-(P'_j-P_j)^2/2\Delta P^2_j]. \label{P'}
\end{equation}
The final score was obtained by adding all the probabilities together

\begin{equation}
{\rm SDC3a_{FOM}}=\sum_j  \Pr(P'_j) \label{score}.
\end{equation}

A possible strategy for solving the challenge is to avoid the most contaminated modes, therefore not submitting all power spectrum entries. 
In the event that $P$, $\Delta P$ are not submitted for some $j$, a broad, pre-defined prior $\tilde{P}\pm \Delta \tilde{P}$ was used. With the spectrum $\tilde{P}$ we wanted to capture the possible sources of error, due to noise and foreground contamination. We defined it as 
\begin{equation}
\tilde{P}=P'+P_{\rm foreg}\times f^2_{\rm foreg}+P_{\rm noise}
\end{equation}
where $P_{\rm foreg}$ and $P_{\rm noise}$ are the cylindrical power spectra of the foregrounds and of the noise, respectively, and $f_{\rm foreg}$ is a scaling factor to somewhat reduce the foreground contamination. We used $f_{\rm foreg}=0.1$, which represents a reduction in the foreground map intensity by one order of magnitude. We further adopted $\Delta \tilde {P}=\tilde{P}$ to reflect a non-detection of this mode. Table \ref{tab:avoidance} lists the teams which employed foreground avoidance as part of their strategy, and the percentage of data entries submitted as a function of frequency.

Table \ref{tab:leaderboad} presents the SDC3a leaderboard based on the SDC3a score. Scores are included for the top 10 teams. The best score is achieved by the team DOTSS-21cm with the Advanced ML-GPR and the ML-GPR foreground removal strategies, followed by HIMALAYA. 

%

%Table \ref{tab:score_freq} shows the percentage of the total score achieved as a function of frequency. Most commonly, the score steadily improves with higher frequency, which is consistent with foreground contamination being stronger at lower frequency. In some cases the trend is very marked, with the lowest frequency channel(s) essentially not contribution to the score. 

%Table \ref{tab:score_kpar} shows the percentage of the total score achieved as a function of $k_\parallel$. 

%Table \ref{tab:score_kper} shows the percentage of the total score achieved as a function of $k_\bot$. 

In the rest of the section we explore a variety of metrics and figures to further discuss the challenge outcome. 
Given the choice of multiple frequency intervals and 2D power spectra as challenge submission, and the high number of teams that completed the challenge, a thorough inspection of the results requires a large number of figures. In the rest of this section we include only some of them, for illustration and discussion purposes. The complete set of figures is made available in an online repository \footnote{\url{https://tinyurl.com/SDC3a}}.  
In all team comparison plots, to reduce figure complexity, we show only the best-scoring entry for the DOTSS-21cm and KUSANAGI teams that submitted multiple results. Whenever results are shown for a single frequency bin rather than the whole 106--196\,MHz range, we include here the 151--166\,MHz bin, while the other ones are available in the online repository. The reason for this choice is that, for most teams, scores steadily improve with higher frequency, as foreground emission gets fainter, such that the central frequencies represent the in-between scenario. However, some teams exhibit different and sometimes much more marked trends with $\nu$, which means the central bin is not a good representation of the overall performance. This is quantified in Table \ref{tab:score_freq}, which shows how the total score is accumulated between the 6 frequency bins for all teams. 

Figure \ref{fig:fullPS} shows the submission and the comparison with the true input at 151--166\,MHz for the top 5 scoring teams. The first column shows attempts made by different teams to recover the true power spectrum in the bottom left panel of Fig. \ref{fig:truth}. The error patterns (see also the third column) are quite different, with some team displaying a stronger trend with $k_\parallel$ than others. Error bars also exhibit a significant variation between teams. 

%{\bf should we show 121-136 like in some of the previous figures? Should we include the other teams as well?}

\subsection{True vs estimated signal} \label{sec:residuals}
In figure \ref{fig:allsub} we show the diagonal terms  of the recovered $P(k_\parallel,k_\bot)$ for the frequency range 151--166\,MHz. For comparison, the true power spectrum (black diamonds) and the data before foreground removal (black dot-dashed line) are also shown. To aid figure clarity, teams' results are divided into three panels based on alphabetical order. 

Lines that appear incomplete indicate a foreground avoidance strategy. The team KUSANAGI performed an aggressive masking which included the power spectrum diagonal, which means their results do not appear in this figure. SROT's entry is not visible as below the lower limit of the plot. 
All teams achieve a suppression of the foreground power by at least one order of magnitude, but typically many more. An excess power suggests an incomplete foreground removal, while a power deficit could indicate that some of the EoR signal has been removed. Since the challenge submission involves both foreground mitigation and power spectrum estimation, however, errors could be introduced by the power spectrum computation as well.  
Five teams (Akashanga, DOTSS-21cm, ERWA, HIMALAYA, REACTOR) get the closest power spectrum recovery at the frequency and scales shown in the figure. 

To investigate errors across the whole sample (all scales, all frequencies submitted), Fig. \ref{fig:biases} represents the distribution of the errors $P-P'$ in terms of the median (diamonds) and the 16\% and 84\% percentiles (horizontal bar) for each team. Positive and negative values correspond to the power spectrum being overestimated and underestimated, respectively. Some distributions are quite asymmetric, which results in an offset between the median and the centre of the bar. %The distance of the points from the $P-P'=0$ line in units of the bar width allows an easy assessment of how systematic the errors are. 

As expected, the largest absolute errors, shown in the top panel of the figure, correspond to the signal being overestimated, which is indicative of a partial foreground removal. The bottom panel presents a zoom-in on those teams that are the closest to the $P-P'=0$ line.  

DOTSS-21cm and HIMALAYA are confirmed as the teams doing the best in terms of recovery of $P'$; ERWA, Hausos, REACTOR, SKACH, SROT and Shuimu-Tianlai also perform relatively well in terms of $P-P'$. No team's result is completely unbiased, which is not surprising when correcting for strong systematic errors such as foreground contamination. Some biases are positive and some are negative, with amplitudes (as measured by the absolute value of the median) ranging from $4.2 \times 10^{-4}$ to $1.7 \times 10^{-3}\,$K$^2h^{-3}$cMpc$^3$ in those eight teams. By considering different teams together, one would be able to significantly reduce any bias, as different teams are biased in different direction. Seven teams (Akashanga, DOTSS-21cm, ERWA, Hausos, HIMALAYA, SKACH, Shuimu-Tianlai) achieve a value of the median fractional error  $-1 < (P-P')/P' < 1$ across all scales and frequencies, although there are significant outliers in some cases.  

%In table $\ref{tab:chisq}$ we rank the teams in terms of the $\chi^2_{P}$ metric of eq. \ref{chisq}. For this metric, we have not included in the ranking the teams that have performed foreground avoidance, as the $\chi^2$-based metrics resulted to be quite sensitive to the value assigned to the prior $\tilde{P}$. This was not the case for the SDC3a score of eq. \ref{score} because the error bar $\Delta \tilde{P}$ was considered together with $\tilde{P}$. 
%DOTSS-21cm and HIMALAYA are confirmed in the top-3 teams for the $\chi^2_P$ metric as well, with low $\chi^2$ values indicative of a good performance.  

%\begin{table}
%\caption{SDC3a teams ranked by the $\chi^2_P$ metric of eq.(\ref{chisq}) applied to the recovery of the true power spectrum. Note that this metric is not computed for the teams that have used foreground avoidance, listed in Table 1.} 
%\label{tab:chisq}
%\begin{center}
%\begin{tabular}{l l l}
%\hline
%Rank&Team&$\chi^2$\\
%\hline
%1&DOTSS-21cm Advanced ML-GPR&3.3\\
%2&HIMALAYA&7.3\\
%3&DOTSS-21cm ML-GPR&8.1\\
%4&Hausos&9.8\\
%5&SROT&11.3\\
%6&REACTOR&13.8\\
%7&SKACH&16.6\\
%8&Shuimu-Tianlai&22.8\\
%9&ERWA&182.2\\
%10&Akashganga&125890\\
%11&Pisano Galaxy Moppers&\\
%12&KORSDC&\\
%13&Foregrounds-FRIENDS&\\
%14&Nottingham-Imperial&\\
%15&Wizards of Oz 3D&\\
%\hline
%\end{tabular}
%\end{center}
%\end{table}

\subsection{True vs estimated error bar}\label{sec:errorbar}
Figure \ref{fig:allerr} shows an assessment of error bar estimation on the same frequency and scales of Fig. \ref{fig:allsub} ($k_\parallel=k_\bot$ elements only), by means of the submitted error bars $\Delta P(k_\parallel,k_\bot)$ normalised by the absolute value of the true error committed  $|P(k_\parallel,k_\bot)-P'(k_\parallel,k_\bot)|$. Values below and above 1 represent an underestimation and an overestimation of the error bars, respectively. 
This figure demonstrates how accurate error estimation is generally an issue for the teams, with the ratio between estimated and true error in some cases exceeding one order of magnitude. 

Figure \ref{fig:errorbars} compares the distributions of $\log(\Delta P/|P-P'|)$ for all teams at all scales and frequencies, in terms of the median (diamonds) and the 16\% and 84\% percentiles (error bar). The logarithmic scale has been necessary to compress the very high dynamic range of this metric.  Values below and above zero correspond to the error bars being under- and over-estimated, respectively. 

While some teams have been conservative with their error bars (e.g., Akashanga, REACTOR), most commonly error bars have been underestimated. Seven teams (DOTSS-21cm, ERWA, HAMSTER, HIMALAYA, KUSANAGI, Pisano Galaxy Moppers, REACTOR) manage to keep $\Delta P/|P-P'|$ within 0.1 and 10. Pisano Galaxy Moppers performs the best overall, with $\Delta P/|P-P'|=0.4$--0.8 and a median of 0.6. 
%In Table \ref{tab:chisq2} we rank the teams in terms of the $\chi^2_{\Delta P}$ metric of eq. (\ref{chisq2}), where again scores are included for the 10 top-scoring teams. 
%The best $\chi^2_{\Delta P}$ metric is significantly higher than the best $\chi^2_P$, which confirms the fact that error estimation has been less successful overall. The best results in terms of this metric are achieved by Pisano Galaxy Moppers, followed by REACTOR and Akashanga. 
%
%\begin{table}
%\caption{SDC3a teams ranked by the $\chi^2_{\Delta P}$ metric of eq.(\ref{chisq2}) applied to the recovery of the error bar. Note that this metric is not computed for the teams that have used foreground avoidance, listed in Table 1.} 
%\label{tab:chisq2}
%\begin{center}
%\begin{tabular}{l l l}
%\hline
%Rank&Team&$\chi^2_{\Delta P}$\\
%\hline
%1&Pisano Galaxy Moppers&263\\
%2&REACTOR&1660\\
%3&Akashganga&1744\\
%4&HIMALAYA&5920\\
%5&ERWA&6721\\
%6&DOTSS-21cm Advanced ML-GPR&7817\\
%7&Foregrounds-FRIENDS&107849\\
%8&Hausos&263784\\
%9&KORSDC&346795\\
%10&DOTSS-21cm ML-GPR&617388\\
%11&Nottingham-Imperial&\\
%12&Shuimu-Tianlai&\\
%13&SKACH&\\
%14&Wizards of Oz 3D&\\
%15&SROT&\\
%\hline
%\end{tabular}
%\end{center}
%\end{table}
The mixed performance in this metric reflects the difficulty of estimating and propagating systematic errors. %An accurate assessment of those errors may require the analysis of multiple sets of simulated data, which was not possible for the teams to achieve within the time-frame of the challenge. Another strategy that can be used in practice is the comparison between results obtained with independent foreground-cleaning pipelines on the same dataset. This highlights the importance of developing multiple foreground-removal pipelines to apply to the real data.  {\bf ref to fig 6. }

\subsection{Reproducibility badges} \label{sec:reproducibility}

During the SDC3a challenges, each participating team is eligible for a reproducibility badge. The badges are designed to recognise efforts by teams to prepare software pipelines that can be reproduced and reused by others. The Software Sustainability Institute (SSI)\footnote{ \url{https://www.software.ac.uk/} } \citep{crouch2013software} recommends six steps towards writing sustainable code\footnote{``How to make your script ready for publication'': \url{https://www.software.ac.uk/guide/how-make-your-script-ready-publication}}, enabling others to use and modify it.  An SDC3a reproducibility badge is be awarded to each team who follows the recommended steps:

\begin{enumerate}
    \item Put code under version control
    \item Make sure your code is in a shareable state
    \item Add essential documentation
    \item Add a license 
    \item Mark the stable version of your code
    \item Make your code citable
\end{enumerate}

Nine teams submitted entries for the SDC3a reproducibility badges.
Each pipeline was evaluated by an expert panel against the pre-defined criteria. Table~\ref{repawards} reports the teams and pipelines to which badges were awarded.

\begin{table*}
	\centering
	\begin{tabular}{lll} % four columns, alignment for each
     %   	\hline\noalign{\smallskip}
       % 	\smallskip
        	 Team name & Pipeline & Reference\\

    	 	 \hline\noalign{\smallskip}
			  Cantabrigians	  &\url{https://github.com/ycliu23/Cambridge-SKA-SDC3-Foregrounds}&\cite{SKAO_Science_Data_Challenge_3a_Liu}\\
           DOTSS &\url{https://gitlab.com/flomertens/dotss21_sdc3_pipeline}	&\cite{mertens_2023_10263162}\\
            ERWA &\url{https://github.com/zzh0616/SKA-DECONV}	&\cite{ERWA_sdc3a}\\
           FOREGROUNDS-FRIENDS	 &\url{https://github.com/espsrc/FOREGROUNDS-FRIENDS}		&\cite{Foregrounds-FRIENDS}\\
            Hausos		&\url{https://github.com/CEA-jiangming/Hausos-sdc3a/tree/v1.0}	&\cite{Hausos_sdc3a}\\
            HIMALAYA	&\url{https://github.com/553445316/HIMALAYA.git}	&\cite{HIMALAYA_sdc3a}\\
            KORSDC		&\url{https://github.com/KJ-Ahn/KORSDC_FGremove}	&\cite{KORSDC_sdc3a}\\
            SROT	&\url{https://github.com/AkashRadio/SKA_SDC3}	&\cite{SROT_sdc3a}\\
           Wizards of Oz	 &\url{https://github.com/d3v-null/sdc3-pipeline}	&\cite{SKAO_Science_Data_Challenge_3a}\\
     	%\hline \noalign{\smallskip}
	\end{tabular}
    	\caption{Reproducibility badges were awarded to nine teams who followed  best practice in the provision of reproducible results and reusable methods. Entries were evaluated using the recommendations of the Software Sustainability Institute (Section~\ref{sec:reproducibility} }
        \label{repawards}
\end{table*}

\section{Discussion and conclusions} \label{sec:conclu}
%The challenges posed by the removal of foreground contamination and the treatment of instrumental systematics have been the limiting factor in the detection of the EoR signal from currently operating experiments \citep{}. 
%Although very challenging in itself, the removal of diffuse foreground contamination as an isolated step of the data analysis pipeline has been performed  successfully in previous work on simulations \citep[e.g.][and references therein]{chapman2015}. However, a lot of additional complexity is introduced when this step is combined with the removal of point-like sources and coupled with instrumental systematics, which is where the challenge with real data lies.

SDC3a aimed at bringing simulation work close to the realistic scenario, by including both diffuse and point-like foreground contamination, as well as the residual effect of out-of-field strong sources that enter the FoV due to the telecope's sidelobes, ionospheric effects, and calibration errors. The present work does not include systematics induced by polarization, for example polarization leakage.  
The increased realism mandates the use of visibility data, rather than image-plane data, for the data simulation and, typically, for part of the data analysis performed by teams. The resulting analysis is more complex and more computationally demanding than some previous exercises, which needs to be taken into account when comparing with previous results. 

There were 20 submissions of results at the end of the challenge, coming from teams all around the world. As in previous SKA SDCs, different teams had different level of expertise in this analysis, and employed pipelines ranging from "tried and tested" to still under development, both of which contribute to the difference in teams' performance. Some results exhibit systematics that suggest an issue with the actual computation of the power spectrum (e.g. the binning in $k$, the correction of the PSF and/or PB) rather than the foreground reduction, which complicates the comparison of the results. 
Several performance metrics are presented in this work for all competing teams, giving feedback and indicating paths for improvement. Additionally, the release of the ground truth after the challenge completion allows teams to further inspect their results. 

All teams achieved a significant suppression of the initial contamination from foreground emission, however in some cases the result is still significantly biased, mostly -- but not exclusively -- in excess of the true signal. 
The best performance in terms of EoR power spectrum recovery, from the team DOTSS-21cm, is $P-P'=4.2_{-4.2}^{+20} \times 10^{-4}$\,K$^2h^{-3}$cMpc$^{3}$ over all frequencies and scales probed by the challenge. For comparison, the true EoR power spectrum across all frequencies and scales is $P'=1.7_{-1.3}^{+14} \times 10^{-3}$\,K$^2h^{-3}$cMpc$^{3}$, which means that the residual error is about four times lower. Six more teams (ERWA, HAMSTER, HIMALAYA, KUSANAGI, Pisano Galaxy Moppers, REACTOR) also achieve a median residual error below the true EoR power spectrum.

The estimation of the error bars $\Delta P$ (approximated as Gaussian and uncorrelated) presented a challenge for the teams, with the true error committed being in some cases orders of magnitude away from the estimation. This indicates the need for robust error estimation strategies, since over- or under-estimated error bars can mean the difference between having a detection of the EoR signal or an upper limit, and can lead to wrong conclusions in terms the inference of the reionization properties of the Universe. The best performance in terms of error estimation, from the team Pisano Galaxy Moppers, is $\Delta P/|P-P'|=0.6_{-0.2}^{+0.2}$. 

The difficulty with estimating errors in foreground removal is that they are mostly systematic errors, stemming from effects that the analysis either does not consider or models only partially. As such, they cannot easily be quantified nor propagated. A possible strategy involves relying on parallel analysis on simulations, where the ground truth is known. The success of this strategy for error estimation relies on the accuracy of the simulation, both in terms of the sky signal and the instrumental response. Another approach is the comparison of results on real data between independent pipelines. Figure \ref{fig:biases} shows a promising convergence of several teams around the truth, which indicates that such an approach would be feasible. 

\section{Data availability statement}
Pipelines used the for this analysis are available at the addresses specified in Sec. \ref{sec:reproducibility}. The simulation data is available as described in \cite{sdc3a1}.

\bibliography{sdc3bcopy}
\bibliographystyle{mn2e_plus_arxiv}

\appendix
\section{Score as a function of scale}
Table \ref{tab:score_kpar} shows the percentage of the total score achieved as a function of $k_\parallel$. 

Table \ref{tab:score_kper} shows the percentage of the total score achieved as a function of $k_\bot$.  

A residual foreground contamination typically creates a strong dependence on $k_\parallel$, with low-$k$ modes being more contaminated that high-$k$ modes. A dependence on $k_\bot$ could indicate some systematics in the correction for the primary beam or the PSF in the power spectrum computation.  

\begin{table*}
\caption{Percentage contribution to the total score for different $k_\parallel$ bins} 
\label{tab:score_kpar}
\begin{center}
\begin{tabular}{l l l l l l l l l l l}
\hline
Team&0.05&0.10&0.15&0.20&0.25&0.30&0.35&0.40&0.45&0.50\\
\hline
Akashganga&0&0&2&4&6&7&6&7&9&53\\
Cantabrigians&-&-&0&0&0&4&20&17&33&23\\
DOTSS-21cm Advanced ML-GPR&7&5&5&7&8&8&9&14&16&17\\
ERWA&0&1&3&5&8&11&13&15&19&21\\
Foregrounds-FRIENDS&99&0&0&0&0&0&0&0&0&0\\
HAMSTER&-&-&0&1&8&10&17&19&18&22\\
Hausos&22&18&24&19&2&1&1&3&2&4\\
HIMALAYA&3&8&8&7&7&6&6&8&19&23\\
KORSDC&0&1&6&8&6&7&12&22&18&16\\
KUSANAGI&-&0&0&0&0&0&11&33&26&28\\
Nottingham-Imperial&100&0&0&0&0&0&0&0&0&0\\
Pisano Galaxy Moppers&3&3&4&5&7&9&11&14&17&21\\
REACTOR&6&6&6&7&8&8&10&12&15&18\\
SKACH&1&1&4&7&7&23&19&17&6&11\\
SROT&10&10&10&10&10&10&10&10&10&10\\
Shuimu-Tianlai&2&3&3&4&7&10&10&14&26&16\\
Wizards of Oz 3D&0&0&1&6&7&9&11&10&13&38\\
\hline
\end{tabular}
\end{center}
\end{table*}

\begin{table*}
\caption{Percentage contribution to the total score for different $k_\bot$ bins} 
\label{tab:score_kper}
\begin{center}
\begin{tabular}{l l l l l l l l l l l}
\hline
Team&0.05&0.10&0.15&0.20&0.25&0.30&0.35&0.40&0.45&0.50\\
\hline
Akashganga&1&0&1&2&5&8&13&17&20&28\\
Cantabrigians&98&1&0&0&0&0&0&0&0&0\\
DOTSS-21cm Advanced ML-GPR&5&8&8&10&10&12&13&12&9&7\\
ERWA&4&5&6&6&8&10&11&13&15&18\\
Foregrounds-FRIENDS&6&93&0&0&0&0&0&0&0&0\\
HAMSTER&2&12&15&10&15&10&6&11&7&7\\
Hausos&5&1&3&3&3&6&9&14&21&31\\
HIMALAYA&4&6&8&10&6&12&11&10&13&15\\
KORSDC&100&0&0&0&0&0&0&0&0&0\\
KUSANAGI&48&27&1&22&0&0&-&-&-&-\\
Nottingham-Imperial&0&0&0&0&0&0&0&2&31&66\\
Pisano Galaxy Moppers&5&5&5&6&7&9&11&13&15&19\\
REACTOR&2&3&4&5&6&8&11&15&18&23\\
SKACH&0&44&22&32&0&0&0&0&0&0\\
SROT&10&10&10&10&10&10&10&10&10&10\\
Shuimu-Tianlai&1&7&10&9&11&11&11&11&12&12\\
Wizards of Oz 3D&4&6&8&13&12&10&11&13&12&6\\

\hline
\end{tabular}
\end{center}
\end{table*}
\clearpage
\onecolumn\section*{Affiliations}
1 SKA Observatory, Jodrell Bank, Lower Whitington, Macclesfield, SK11 9FT, UK\\2 Max-Planck Institute for Astrophysics, Karl-Schwarzschild-Stra{\ss}e 1, 85748 Garching, Germany\\3 Department of Earth Sciences, Chosun University, Gwangju 61452, Korea\\4 Universidad Europea de Madrid, 28670, Madrid, Spain\\5 Laboratoire d’Astrophysique, Ecole Polytechnique Federale de Lausanne (EPFL), Observatoire de Sauverny, Versoix 1290, Switzerland\\6 Department of Physics and Trottier Space Institute,  McGill University, 3600 rue University, Montréal, QC H3A 2T8, Canada\\7 School of Physics and Astronomy, The University of Nottingham, University Park, Nottingham, NG7 2RD\\8 Department of Physics and Astronomy, University of the Western Cape, 7535 Bellville, Cape Town, South Africa\\9 Kapteyn Astronomical Institute, University of Groningen, PO Box 800, 9700 AV Groningen, The Netherlands\\10 Research Center for Astronomical Computing, Zhejiang Laboratory, Hangzhou 311121, China\\11 National Astronomical Observatories, Chinese Academy of Sciences, Beijing 100101, China\\12 Institute for Astronomy, The University of Edinburgh, Royal Observatory, Edinburgh EH9 3HJ, UK\\13 Instituto de Física de Cantabria (CSIC-Universidad de Cantabria), Avda. de los Castros s/n, E-39005 Santander, Spain\\14 Instituto de Astrofísica de Andalucía (IAA-CSIC), Glorieta de la Astronomía s/n, 18008, Granada, Spain\\15 Haute Ecole Arc Ing\'enierie, University of Applied Sciences and Arts Western Switzerland (HES-SO), Saint-Imier, Switzerland\\16 Centre for Artificial Intelligence, ZHAW Zurich University of Applied Sciences, Technikumstrasse 71, 8400 Winterthur, Switzerland\\17 Department of Astronomy, Tsinghua University, Beijing 100084, China\\18 D\'epartement de Physique Th\'eorique and Center for Astroparticle Physics, Universit\'e de Geneve, 24 quai Ernest Ansermet, 1211 Geneve 4, Switzerland\\19 Nordita, KTH Royal Institute of Technology and Stockholm University, Hannes Alf\'vens v\"ag 12, SE-106 91 Stockholm, Sweden\\20 Korea Astronomy and Space Science Institute, Daejeon 34055, Korea\\21 Astronomy Campus, University of Science and Technology, Daejeon 34055, Korea\\22 Kumamoto University, International Research Organization for Advanced Science and Technology, 2-39-1 Kurokami, Chuo-ku, Kumamoto 860-8555, Japan\\23 National Key Laboratory of Radar Signal Processing, Xidian Universtiy, Xi'an, 710071, China\\24 Guangzhou Institute of Technology, Xidian University, Guangzhou, 510555, China\\25 International Centre for Radio Astronomy Research, Curtin University,  Bentley WA, Australia\\26 ARC Centre of Excellence for All-Sky Astrophysics in 3D, Australia\\27 ICRAR M468, The University of Western Australia, Western Australia, 6009, Australia\\28 Department of Physics and Astronomy, Sejong University, Seoul 05006, Korea\\29 Korea AeroSpace Administration, Sacheon-si 52535, Gyeongsangnam-do, Korea\\30 Indian institute of Tecnology Dharwad, Karnataka, India 580011\\31 Aurora Technology for the European Space Agency, Camino bajo del Castillo, s/n, Urbanizaci\'on\\32 Villafranca del Castillo, Villanueva de la Canada, Madrid, Spain\\33 Department of Physics, Ulsan National Institute of Science and Technology, Ulsan 44919, Korea\\34 Cavendish Astrophysics, University of Cambridge, JJ Thomson Avenue, Cambridge CB3 0HE, UK\\35 Kavli Institute for Cosmology, University of Cambridge, Madingley Road, Cambridge CB3 0HA, UK\\36 Jodrell Bank Centre for Astrophysics, Department of Physics and Astronomy, The University of Manchester, Manchester M13 9PL, UK\\37 LUX, Observatoire de Paris, PSL Research University, CNRS, Sorbonne Universit\'e, F-75014 Paris, France\\38 Scuola Normale Superiore, Piazza dei Cavalieri 7, I-56126 Pisa, Italy\\39 Blackett Laboratory, Imperial College London, Prince Consort Road, London, SW7 2AZ, UK\\40 Australian SKA Regional Centre (AusSRC), Curtin University, Bentley, WA, Australia\\41 ASTRON Netherlands Institute for Radio Astronomy, Oude Hoogeveensedijk 4, 7991 PD Dwingeloo, The Netherlands\\42 Dpto. de Física Moderna, Universidad de Cantabria, Avda. los Castros s/n, E-39005 Santander, Spain\\43 Shanghai Astronomical Observatory, Chinese Academy of Sciences, 80 Nandan Road, Shanghai 200030, P. R. China\\44 School of Astronomy and Space Science, University of Chinese Academy of Sciences, Beijing 100049, P. R. China\\45 Key Laboratory of Radio Astronomy and Technology, Chinese Academy of Sciences, A20 Datun Road, Chaoyang District, Beijing 100101, P. R. China\\46 Space, Planetary \& Astronomical Sciences \& Engineering (SPASE), Indian Institute of Technology, Kanpur, 208016, Uttar Pradesh, India\\47 Mizusawa VLBI Observatory, National Astronomical Observatory Japan, 2-21-1 Osawa, Mitaka, Tokyo 181-8588, Japan\\48 School of Physics and Astronomy, Sun Yat-sen University, Zhuhai 519082, People’s Republic of China\\49 Liaoning Key Laboratory of Cosmology and Astrophysics, College of Sciences, Northeastern University, Shenyang 110819, China\\50 Laboratory for Advanced Computing of the University of Coimbra and CFisUC, Departamento de Física, Universidade de Coimbra, 3000-056 Coimbra, Portugal\\51 Laboratoire de Physique de l'Ecole Normale Sup\'erieure, ENS, Universit\'e PSL, CNRS, Sorbonne Universit\'e, Universit\'e de Paris, F-75005 Paris, France\\52 KLE Tecnological University, Hubli,Karnataka, India 580031\\53 Research Computing Services, Roger Needham Building, 7 J J Thomson Avenue, West Cambridge Site, Cambridge, CB3 0RB, UK\\54 INAF-Istituto di Radioastronomia,  via P. Gobetti 101, 40129 Bologna, Italy\\55 Centre for Strings, Gravitation and Cosmology, Department of Physics, Indian Institute of Technology Madras, Chennai 600036, India\\56 Center of Fundamental Physics of the Universe, Dept of Physics, Brown University, Providence 02912, Rhode Island, United States\\57 IFREMER, Univ. Brest, CNES, CNRS, IRD, Laboratoire d'océanographie physique et spatiale , France\\58 School of Physics and Astronomy, Sun Yat-Sen University, No.2 Daxue Road, Zhuhai 519082, China\\59 Swiss National Supercomputing Center, ETH Zurich, 6900 Lugano, Switzerland\\60 Fundacion Centro Tecnologico de Supercomputación de Galicia (CESGA), Avda. de Vigo s/n. Campus Sur. 15705 Santiago de Compostela, Spain\\61 INAF-Italian Centre for Astronomical Archives, via G.B. Tiepolo 11, 34143 Trieste, Italy\\62 Computer Vision and Pattern Recognition Unit, Indian Statistical Institute, Kolkata, India\\63 Département de physique, de génie physique et d’optique, Université Laval, Québec (QC), G1V 0A6, Canada\\64 Suzuka National College of Technology Department of Mechanical Engineering, Shiroko-cho, Suzuka, Mie, 510-0294, Japan\\65 National Astronomical Observatory of Japan, 2-21-1 Osawa, Mitaka, Tokyo 181-8588, Japan\\66 Azabu Junior and Senior High School, 2-3-29 Motoazabu, Minato, Tokyo 106-0046, Japan\\67 Ru{\dj}er Bo\v{s}kovi\'c Institute, Bijeni\v{c}ka cesta 54, 10000 Zagreb, Croatia\\68 IDRIS, CNRS, Université Paris-Saclay, 91403, Orsay, France\\69 School of Physics and Electronic Science, Guizhou Normal University, Guiyang 550001, China\\70 Department of Physics, National Institute of Technology Calicut, Calicut 673601, Kerala, India\\71 Department of Astronomy, Astrophysics, and Space Engineering, Indian Institute of Technology Indore, Madhya
Pradesh, 453552, India\\72 School of Earth and Space Exploration, Arizona State University, Tempe, AZ 85287-1404, USA\\73 Indian Institute of Science\\74 School of Physics, The University of Melbourne, Victoria, Australia\\75 Instituto de Astrofisica de Canarias, E-38205 La Laguna, Tenerife, Spain\\76 Departamento de Astrofísica, Universidad de La Laguna, E-38206 La Laguna, Tenerife, Spain\\77 Physics and Applied Mathematics Unit, Indian Statistical Institute, 203 B.T. Road, Kolkata 700 108, India\\78 Department of Computer Science, University of Nevada Las Vegas, 4505 S. Maryland Pkwy., Las Vegas, NV 89154, USA\\79 School of Aerospace Science and Technology, Xidian University, Xi'an, 710126, China\\80 Yunnan University, SWIFAR, No. 2 North Green Lake Road, Kunming, Yunnan Province 650500, China\\81 Indian Institute of Technology, Kanpur\\82 Université Paris-Saclay, Université Paris Cité, CEA, CNRS, AIM, 91191, Gif-sur-Yvette, France\\83 Institutes of Computer Science and Astrophysics, Foundation for Research and Technology Hellas (FORTH), GR 70013 Heraklion, Crete, Greece\\84 Division of Particle and Astrophysical Science, Nagoya University, Furo-cho, Chikusa-ku, Nagoya, 464–8602, Japan\\85 The Research Center for Statistical Machine Learning, The Institute of Statistical Mathematics, 10-3 Midori-cho, Tachikawa, Tokyo 190–8562, Japan\\86 INAF - Osservatorio Astrofisico di Catania,  via Santa Sofia 78, 95123 Catania, Italy\\87 School of Physics and Astronomy, Beijing Normal University, Beijing 100875, China\\88 Department of Physics, Stellenbosch University, Matieland 7602, South Africa\\89 National Institute for Theoretical and Computational Sciences (NITheCS), South Africa\\90 School of Artificial Intelligence, Beijing Normal University, Beijing 100875, China\\91 School of Computer Science and Engineering, Xi'an University of Technology, Xi'an, 710048, China\\92 College of Big Data and Information Engineering, Guizhou University, Guiyang 550025, China\\93 State Key Laboratory of Public Big Data, Guizhou University, Guiyang 550025, China\\
\end{document}